\documentclass[journal]{IEEEtran}

\usepackage{caption}
\usepackage{algorithmic}
\usepackage[linesnumbered,ruled]{algorithm2e}
\usepackage{graphics,epstopdf}
\usepackage{balance}
\usepackage{listings}
\usepackage{comment}
\usepackage{enumitem}
\usepackage{fixmath}
\usepackage{cite}
\usepackage{amsmath,amssymb,amsfonts}
\usepackage{graphicx}
\usepackage{textcomp}
\usepackage{tabularx}
\usepackage{booktabs}
\usepackage{longtable}
\usepackage{multirow}
\usepackage{colortbl}
\usepackage{hhline}
\usepackage{tabularx,colortbl}
\usepackage{stmaryrd}
\usepackage{cite}
\usepackage{enumitem}
\usepackage{soul}
\usepackage{bm} %bold math \bm{s}
\usepackage{mathtools} % text on arrow
\usepackage{mathrsfs} % for A as adversary
\usepackage{multirow}
\usepackage{array}
\usepackage{colortbl} 
\usepackage{multirow, booktabs}
\usepackage{hhline, boldline, makecell}

\usepackage{soul}%highlight
\usepackage{multirow}
\usepackage{array, makecell}
\usepackage{diagbox}%
\renewcommand*{\hl}{}
\usepackage{amsmath} % for "\dfrac" macro
\usepackage{ragged2e}
\usepackage{array,makecell,booktabs}
\newcolumntype{M}[1]{>{\centering\arraybackslash}m{#1}}
\usepackage{cellspace}
\usepackage[table,xcdraw]{xcolor}
\usepackage{tikz}

\newcommand*\circleded[1]{\tikz[baseline=(char.base)]{
            \node[shape=circle,draw,inner sep=0.01pt] (char) {#1};}}

\usepackage{enumitem}

\ifCLASSINFOpdf

\else

\fi

\begin{document}

\title{Efficient and Privacy-Preserving Infection Control System for Covid-19-Like Pandemics using Blockchain}

\author{\IEEEauthorblockN{Seham A. Alansari, Mahmoud M. Badr,
Mohamed Mahmoud,~\IEEEmembership{Senior~Member,~IEEE,}\\ and
Waleed Alasmary,~\IEEEmembership{Senior~Member,~IEEE}  \vspace{-1cm}\\}
\thanks{Corresponding author: Mohamed Mahmoud.}
\thanks{S. A. Alansari and W. Alasmary are with the Department of Computer Engineering, Umm Al-Qura University, Saudi Arabia (e-mail: s43980120@st.uqu.edu.sa; wsasmary@uqu.edu.sa).}
\thanks{M. M. Badr and M. Mahmoud are with the Department of Electrical and Computer Engineering, Tennessee Tech. University, Cookeville, TN 38505 USA (e-mail: mmbadr42@tntech.edu; miibrahem42@tntech.edu; mmahmoud@tntech.edu).}
\thanks{
Copyright (c) 20xx IEEE. Personal use of this material is permitted. However, permission to use this material for any other purposes must be obtained from the IEEE by sending a request to pubs-permissions@ieee.org.
} 
}

\maketitle
%\IEEEdisplaynontitleabstractindextext

% The paper headers
\markboth{Seham \MakeLowercase{\textit{et al.}}: Efficient and Privacy-Preserving Infection Control System for Covid-19-Like Pandemics using Blockchain}%
{}
\IEEEpeerreviewmaketitle
%\IEEEtitleabstractindextext{%
\begin{abstract}
Contact tracing is a very effective way to control the COVID-19-like pandemics. 
It aims to identify individuals who closely contacted an infected person during the incubation period of the virus and notify them to quarantine. 
However, the existing systems suffer from privacy, security, and efficiency issues. 
To address these limitations, in this paper, we propose an efficient and privacy-preserving Blockchain-based infection control system. 
Instead of depending on a single authority to run the system, a group of health authorities, that form a consortium Blockchain, run our system. 
Using Blockchain technology not only secures our system against single point of failure and denial of service attacks, but also brings transparency because all transactions can be validated by different parties. 
Although contact tracing is important, it is not enough to effectively control an infection. 
Thus, unlike most of the existing systems that focus only on contact tracing, our system consists of three integrated subsystems, including contact tracing, public places access control, and safe-places recommendation. 
The access control subsystem prevents infected people from visiting public places to prevent spreading the virus, and the recommendation subsystem categorizes zones based on the infection level so that people can avoid visiting contaminated zones. Our analysis demonstrates that our system is secure and preserves the privacy of the users against identification, social graph disclosure, and tracking attacks, while thwarting false reporting (or panic) attacks. 
Moreover, our extensive performance evaluations demonstrate the scalability of our system (which is desirable in pandemics) due to its low communication, computation, and storage overheads.

\end{abstract}

%}

%\titlepgskip=-15pt

\section{Introduction} 
\label{chap:Introduction} 

\begin{comment}
points need to be included in the introduction:
paragraph1 about COVID-19
paragraph2 regarding its' impact on health, society and economy
paragraph3 about countermeasures and contact tracing as one of the solutions been proposed to fight against the spread of covid-19
paragraph4 
paragraph5
\end{comment}
%
COVID-19 virus has spread quickly to all countries in the world causing a large number of infected people and deaths.
By January 2021, the number of infected cases worldwide has reached 86+ million cases, and the number of deaths is around 2 million \cite{WhoDashboard}. 
The devastating impacts of the COVID-19 pandemic on health, society, and world economy exceeded any other crisis since World War II, as the Secretary-General of the United Nations announced. For the impact on the healthcare sector, COVID-19 patients overwhelmed the hospitals and doctors had to make tough life and death decisions in some countries \cite{onder2020case}. 
For the impact on society, the panic and the lockdown caused by COVID-19 outbreak resulted in adverse mental health consequences and common psychological reactions such as anxiety, depression, stress, and lack of sleep \cite{rajkumar2020covid}. 
For the impact on the economy, COVID-19 disrupted many industries and sectors due to sickening workers and the restrictive measures imposed by the governments on the travel and the gathering of the people \cite{ozili2020spillover}. 
For instance, the travel industry and tourism suffered from hefty losses because many countries suspended all air flights and closed their airports. %And, some airlines have been forced to temporarily hold off their operations. 
The aviation industry lost 113 billion dollars in total\cite{ozili2020spillover}.

The high infection level of the virus and the severe symptoms it may cause forced many countries worldwide to impose harsh measures to slow down the spread of the virus.
Some of these measures include isolation of infected people, enforcing partial/complete lockdown of cities and provinces \cite{9321817}, promoting social-distancing \cite{9138412}, preventing large gatherings, and advising citizens to wear masks. 
However, an individual may be contagious but does not show any symptoms during the incubation period of the virus, which is 14 days in case of COVID-19.
During this period, he may interact with people and pass the virus to them. 
Therefore, contact tracing is one of the most effective ways to control COVID-19-like pandemics \cite{9086010}. 
%employed by various countries.
It aims to identify individuals who closely contacted an infected individual during the incubation period of the virus.
The health authorities in Germany performed manual contact tracing in an attempt to slow down the spread of COVID-19, which showed success in the first few weeks, but manual tracing was not practical when the number of cases increased \cite{Spiegel}. 
This indicates that contact tracing can be effective to contain COVID-19 like pandemics, but it needs to be automated to deal with the large number of infected cases.

Multiple automated contact tracing systems \cite{Tracetogether,Aarogya,wancontactchaser,altuwaiyan2018epic,jhanwarphyct,brack2020decentralized,torky2020covid,arifeen2020Blockchain, reichert2020privacy,troncoso2020decentralized} have been developed by governmental and private institutions. 
These systems depend on exchanging messages between the users' smartphones when they encounter each other as a proof of contact. 
Then, if a user is tested positive for COVID-19, the messages collected by the user during the incubation period are used to identify the close contacts, and thus these people are notified to quarantine to prevent the spread of the virus. 
%One way to do that is by collecting the messages from all infected users 
%by a central authority that notifies the close contacts. 
However, the existing contact tracing systems suffer from privacy, security, and efficiency issues. 

%However, the existing systems lack privacy and security or  have not been considered as one of the prime objectives.

%The aforementioned contact tracing systems in the literature are vulnerable to one or multiple privacy attacks: identification attack (i.e., linkage of the received contact messages to identify a particular person or try to keep track of users' health status whether if he/she is diagnosed as a carrier of the virus), tracking attack (i.e., try to track an individual based on location data) and social graph disclosure attack  (i.e., try to identify contact relationship). Furthermore, these schemes have limitations in  computation, storage usage, and/or scalability. Some of these systems \cite{Tracetogether,Aarogya,wancontactchaser,altuwaiyan2018epic,jhanwarphyct,reichert2020privacy} are based on fully/partially centralized architecture that is subjected to a single point of failure problem and Denial of Service (DoS) attack, which could affect the system's availability or reliability. Some of these schemes suffer from panic/false reporting attack (i.e., make health authority to falsely believe an individual is one of the close contacts of someone infected). %\cite{Tracetogether,Aarogya,wancontactchaser,altuwaiyan2018epic,jhanwarphyct,reichert2020privacy,torky2020covid,brack2020decentralized} 

In terms of privacy, the existing systems are vulnerable to one or multiple privacy attacks such as identification, tracking, and social graph disclosure. 
In identification attack, an adversary tries to identify a particular person from his messages transmitted in the system and learn his COVID-19 status, i.e., whether he/she is infected with COVID-19.
In tracking attack, an adversary tries to learn the locations visited by a person using the messages collected by the system. 
In social graph disclosure attack, an adversary tries to learn social relationships, i.e., the users who met the infected user. 
\textit{Without privacy preservation, the system can be misused to spy on the people} to learn their visited locations and social activities, which may discourage the users to use the system.
In terms of security, most of the existing systems \cite{Tracetogether,Aarogya,wancontactchaser,altuwaiyan2018epic,jhanwarphyct,reichert2020privacy} are based on centralized architecture which is vulnerable to a single point of failure problem and denial of service (DoS) attack, and the lack of transparency. 
%These problems could affect the system's availability and reliability. 
Furthermore, some of the existing systems suffer from false reporting (or panic) attack, where an adversary can deceive the system to incorrectly classify individuals as close contacts.
In terms of efficiency, the existing systems suffer from large communication and computation overheads and inefficient storage usage, especially when the number of users increases.

To address the above limitations, in this paper, we propose an efficient and privacy-preserving Blockchain-based infection control system. 
Instead of depending on a centralized authority to run the system, a group of health authorities, that form a consortium Blockchain, runs our system. 
Using Blockchain technology not only secures our system against the problems caused by the centralized architecture, but also brings transparency because all transactions can be validated by different parties.
Although contact tracing is important, it is not enough to effectively control the infection. Thus, unlike most of the existing systems \cite{Tracetogether,Aarogya,wancontactchaser,altuwaiyan2018epic,jhanwarphyct, reichert2020privacy,troncoso2020decentralized,brack2020decentralized,torky2020covid} that focus only on contact tracing, our system consists of three integrated subsystems, including \textit{contact tracing}, \textit{public places access control}, and \textit{safe-places recommendation}.
% to protect people from getting infected while proceeding with their daily life activities, trace and notify them if they have been close to someone who turns to be infected with COVID-19. Also, it aims to restrict the movement of COVID-19 confirmed cases 
% %\textcolor{red}{who continue their recovery at home and should follow the quarantine instructions and pursuit self-isolation} 
% until their test results become negative. 

The \textit{contact tracing} subsystem is responsible for identifying the individuals who closely contacted an infected person during the incubation period of the virus and notifying them to quarantine.
The \textit{access control} subsystem is responsible for providing a digital pass to users to prevent infected users from accessing public places, such as restaurants, institutions, libraries,...etc., to prevent spreading the virus.
%Non-infected individuals (i.e., their COVID-19 status is negative) can use this tool to proceed with their daily activities and visit different places 
%While the infected individuals and their close contacts (i.e., their COVID-19 status is positive or suspected) are required to isolate themselves from society to stop any further transmission of the virus. 
The \textit{recommendation} subsystem is responsible for categorizing zones based on their infection level. 
Users can use this information to avoid visiting contaminated zones to prevent spreading the virus.
Our security and privacy analysis demonstrate that our system is secure and preserves the privacy of the users against identification, social graph disclosure, and tracking attacks, while thwarting the false reporting attack. 
Moreover, our extensive performance evaluations demonstrate the scalability of our system due to its low communication, computation, and storage overheads.

Our main contributions can be summarized as follows:
%a privacy-preserving proximity detection scheme that can execute contact tracing, effectively help identify and notify the close contacts of the infected cases, and protect non-infected individuals from visiting COVID-19 hotspot zones while allowing them to continue their daily life activities.  we propose a privacy-preserving contact tracing system based on Blockchain technology to protect people from getting infected while proceeding their daily life activities and notify them if they have been close to someone who turns to be infected with COVID-19.
% need to rewrite following points
\begin{itemize}

\item Our system consists of three integrated subsystems including \textit{contact tracing},  \textit{public places access control}, and \textit{safe-places recommendation}, while most of the existing systems focus only on contact tracing which is not enough to control the infection.

\item Our system is secure against multiple privacy attacks including, identification, social graph disclosure, and tracking attacks. Moreover, it is secure against false reporting (or panic) attack.
\item Our system is scalable because it is efficient in terms of computation and communication, and also requires low storage space. %due to the utilization of Bloom filters.
\end{itemize}
%\subsection{Thesis Outline\label{sec:Thesis-Outline}}

The remainder of this paper is organized as follows. 
Section~\ref{chap:preliminaries} covers preliminaries and essential background. The network and threat models and design objectives are explained in section~\ref{chap:network-model}. 
The details of the proposed infection control system are presented
 in section~\ref{chap:proposed scheme}.
%\hl{The proposed scheme is presented
%in details in section}~\ref{chap:proposed scheme}.
The security and privacy analysis are provided in section  \ref{chap:securityandprivacy}, and the performance evaluations are provided in section~\ref{evaluate}. Section~\ref{chap:Literature Review} discusses the related works. Finally, conclusions are drawn in section~\ref{chap:conclusion}.
%\end{comment}
%\end{comment}

\section{Preliminaries}
\label{chap:preliminaries} 

In this section, we present the necessary background on Blockchain, bilinear pairing, and Bloom filter. 

\subsection{Blockchain }\label{sec:Blockchain}

Blockchain is a distributed ledger that consists of blocks of data chained together using cryptographic hash functions. 
The content of the ledger is shared by the entities of a peer-to-peer network and they agree on the content of the ledger using a consensus algorithm \cite{8642861}. 
Blockchain allows entities that do not trust each other to transact without the reliance on a centralized trusted entity \cite{8642861}. 
Blockchain was initially proposed as an enabling technology for cryptocurrencies, but later it has been used to secure many other applications and build secure systems \cite{8642861}. 
The decentralized network architecture of Blockchain secures systems against the single point of failure problem and the DoS attacks. 
Moreover, Blockchain provides immutability in the sense that once a data is written in the ledger, it is impossible to maliciously modify it because this requires controlling the majority of the network nodes, which is impossible to achieve. 
Blockchain also provides transparency because all the transactions are validated by all the network nodes \cite{badr2020smart}.

There are two types of Blockchains, namely permissionless and permissioned \cite{8029379}. 
Permissionless Blockchain is a public Blockchain, where any entity can join the network and all the network entities can read/write from/on the Blockchain \cite{8029379}. 
On the other hand, in permissioned Blockchain, the network access is limited only to authorized entities. One type of permissioned Blockchains is the consortium Blockchain in which only a group of entities have the right to write on the Blockchain and all entities have the right to read from the Blockchain \cite{8029379}. 
% \hl{Generally, there are two broad categories of the Blockchain} \cite{8029379}, \hl{a permissionless Blockchain, where any one can join the Blockchain network and all the users have the same read/write access rights, and a permissioned Blockchain, where the read/write access rights are permitted to authorized users. In this paper, we use a consortium Blockchain} \cite{8029379}\hl{, which is a type of the permissioned blockchians, where only a group of authorized (certified) nodes have the write permission on the shared ledger, and the other users can only query the Blockchain to read data from the ledger.
Consortium Blockchain is appropriate for our system, where  
only the health authorities are able to do COVID-19 tests and they should be able to update the status of the users on the Blockchain, while, the users need only to check their status by reading data from the Blockchain.
% this is suitable for our parking system because only a group of users (the parking lots in our system) have parking offers and need to write them on the shared ledger so they act as the Blockchain validators, while the other users (the drivers in our system) need to query the Blockchain to retrieve the parking offers.

The group of entities that have the right to write on the Blockchain are called validators. 
To maintain a unified ledger, validators have to agree on the content of the ledger. 
This is done through using consensus algorithms. The existing consensus algorithms can be classified into proof-based and voting-based \cite{nguyen2018survey}. One type of voting-based algorithms is Raft in which one validator is elected every period as a leader to add a new block to the ledger \cite{ongaro2014search}. In our system, we use the Raft consensus algorithm because it is more efficient than proof-based consensus algorithms such as proof of work or proof of stake. In Raft, one validator acts as a leader and the other validators act as followers. Every election period, any follower who wants to be a leader becomes a candidate and collects votes from the followers. Finally, the candidate who collects the majority of votes becomes the leader. It is noteworthy to mention that Raft is used in several applications such as Quorum Blockchain of JPMorgan system \cite{rafota}. For more details about Blockchains and consensus algorithms, we refer to \cite{8716424, 8029379, nguyen2018survey, ongaro2014search}.
% A validator in Raft can have one of three different roles: follower, candidate, or leader. Initially, all the validators start as followers. Then, any follower who intends to become a leader should transfer to the candidate state for a period of time (election period) in order to collect enough votes to become a leader. During the election period, all the follower nodes vote for one of the candidates. After the election period timeout, the candidate which collects more votes becomes a leader. The Raft consensus algorithm has been used in Quorum Blockchain of JPMorgan system \cite{rafota}. We refer to \cite{8716424, 8029379, nguyen2018survey, ongaro2014search} for more details on the consortium Blockchain and the Raft consensus algorithm. 
%%%%%%%%%%%%%%%%%%%%%%%%%%%%%%%%%%%%%%%%%%%%%%%%%%%%%%%%%%%%%%%%%%%%%%%%%%%%%%%%%%%%%%%%%%%%%%%%%%%%%%%
\subsection{Bilinear Pairing}
Let $G$ be a cyclic additive group of prime order $q$ and has a generator $P$, and $G_T$ be a cyclic multiplicative group of the same prime order $q$. Let $e$: $G \times G$ $\rightarrow$ $G_T$ be a bilinear map with the following properties. 
\begin{itemize}
\item $e(aP, bQ$)  = $e(P,Q)^{ab}$, where $P, Q \in$ $G$, and $a$, $b$ $\in$ $Z_q$. 
\item $e(P+P_1, Q$)  = $e(P,Q) e(P_1,Q)$, where $P, P_1, Q \in$ $G$.
\item There exists $P,Q$ $\in$ $G$ such that $e(P,Q$) $\neq$ 1. 
\item There is an efficient algorithm to compute $e$($P,Q$) for all $P,Q$ $\in$ $G$.
\end{itemize}
%%%%%%%%%%%%%%%%%%%%%%%%%%%%%%%%%%%%%%%%%%%%%%%%%%%%%%%%%%%%%%%%%%%%%%%%%%%%%%%%%%%%%%%%%%%%%%%%%%%%%%%
\subsection{Bloom Filter} \label{BloomFilter}
\begin{figure}[!t]
    \centering
    \includegraphics[width=0.9\columnwidth]{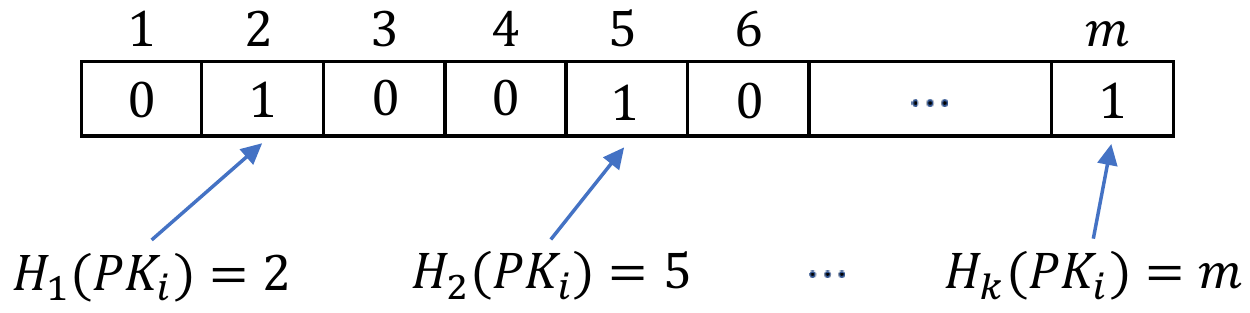}
    \caption{An illustration for the Bloom filter.}
    \label{Bloom}
\end{figure}
Bloom filter is an efficient data structure that has been used in different applications \cite{4016143}, In particular, the filter is a bit vector used to store a list of elements in an efficient way. Fig. \ref{Bloom} illustrates the basic idea of Bloom filter. 
The Bloom filter in Fig. \ref{Bloom} consists of $m$ bits that are initially set to zeros \cite{4016143}.
To store an element (e.g., $PK_{i}$) in the filter, $k$ different hash functions ($H_{1}, H_{2}, \ldots, H_{k}$) are used to calculate the $k$ hash values of the element. 
The output of each hash function gives a number between $1$ and $m$ \cite{7192615}, i.e., points at a bit position in the vector. 
To add an element to the list, the bits the hash functions point at are set to ones.
% The filter is a data structure (or bit array) that stores a set of elements compactly in a bit-vector $B=\left\{B_{1}, B_{2}, \ldots, B_{m}\right\}$ with length of $\mathrm{m}$ bits [16]. Initially, all the vector's $m$ bits are zeroes.Each certificate's public key $\left(PK_{i}\right)$ to be stored should be hashed with $K$ independent hash functions, $\left\{H_{1}(), H_{2}(), \ldots, H_{K}()\right\}$ Each hash function maps a given certificate public key to a number in the range $\{0,1, \ldots, m-1\}$ which represents the address of a bit location in $B$. The result of hashing $PK_{i}$ $K$ times is a sequence of $K$ hash values $\left\{n_{1}, n_{2}, \ldots, n_{k}\right\} \in$ $\{0,1, \ldots, m-1\} .$ To store an element $PK_{i},$ the bits that are pointed by the addresses resulted from the $K$ hashes should be set.
To check if an element is stored in the filter, the $k$ hash values of this element are calculated. 
If all the $k$ locations of the hash values store ones, the element is likely stored in the filter; otherwise, the element is not definitely stored in the filter.
It is possible that an element is not stored in the filter but all the locations resulted from the $k$ hash functions are ones because by coincidence the ones in the $k$ locations come from other elements, and this case is called false positive. The probability of false positive is calculated as follows \cite{851975}.
\begin{equation}
\label{falsepropab}
    \left(1-(1-1 / m)^{k n}\right)^{k}
\end{equation}
where $n$ is the number of elements stored in the filter. By properly selecting the parameters of this equation, the false positive probability can be small.

\section{NETWORK/THREAT MODELS AND DESIGN GOALS}
\label{chap:network-model} 
In this section, we explain the network and threat models considered in this paper and the design goals of our system.

\subsection{Network Model}
\begin{figure*}[!t]
\center \includegraphics[width=7in]{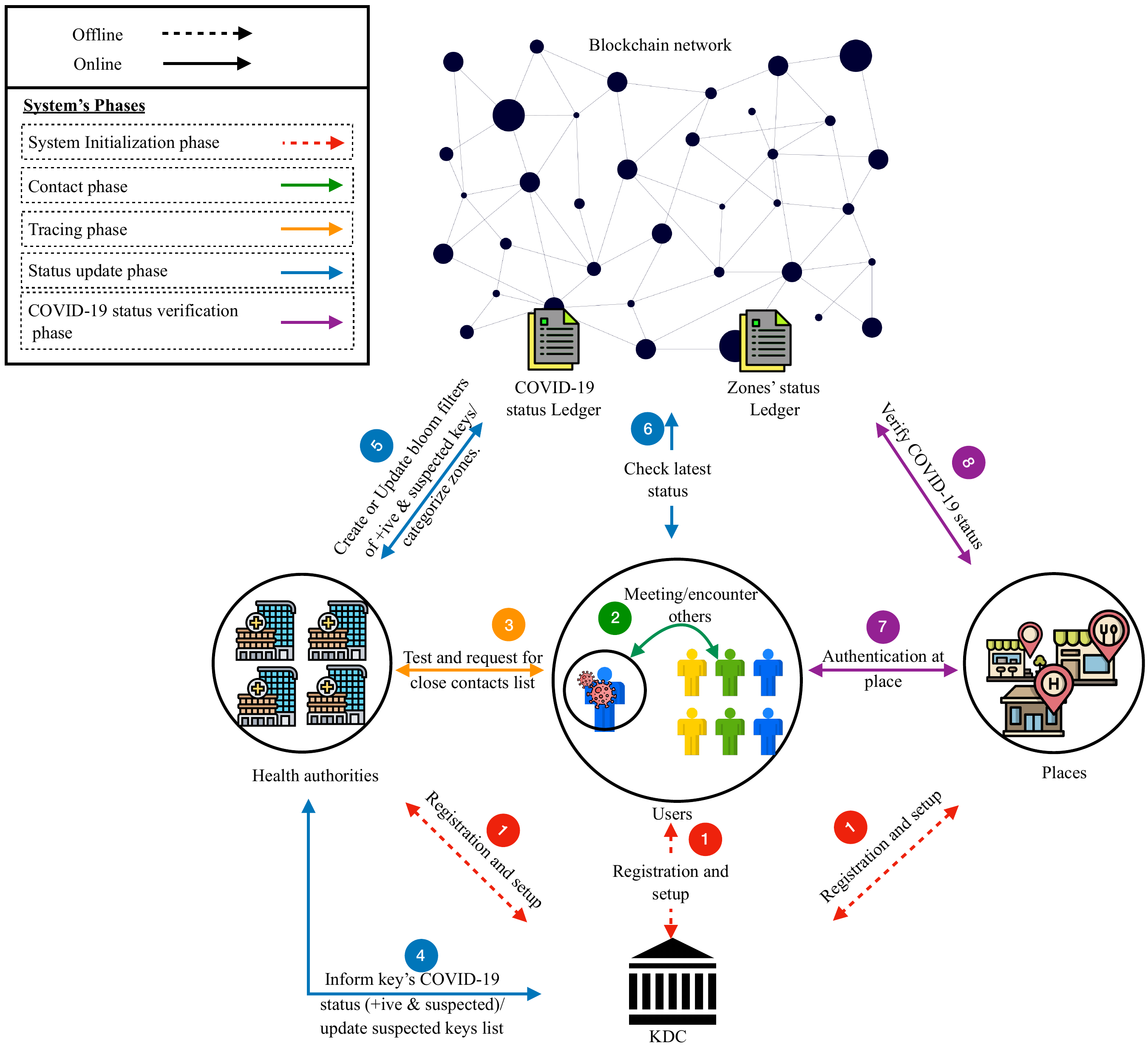}
\caption{An illustration for the network model.} %\textit{Initialization } phase.} 
\label{fig:NetworkModel}
\end{figure*}

As illustrated in Fig.~\ref{fig:NetworkModel}, the network model considered in this paper has five main entities, including \textit{a key distribution center (KDC)}, a \textit{consortium Blockchain}, \textit{health authorities}, \textit{users}, and \textit{public places}. 
The role of each of these entities and the type of the communications among them are as follows.  

%\begin{itemize}

%\item
\textbf{KDC.} The KDC is responsible for initializing the entire system.
This includes the registration of users, health authorities, and places, and also the generation and renewal of their cryptographic keys. 
%\textcolor{red}{based on the proofs obtained from the health authorities in case if they were infected or suspected of having COVID-19}.
To preserve privacy, each user receives anonymous credentials that are used only for a short time.
Moreover, as shown in the figure, the health authorities need to communicate with the KDC, so that it does not renew the anonymous credentials of the infected users.

%it informs the health authorities with the latest keys of the list of suspected users (close contacts) to be used in the creation of a new Bloom filter containing valid keys of suspected users. 

%The KDC is just considered as a system initializer; thus, it does not conflict with the system's decentralization feature. In reality, the KDC can be the ministry of health.

%The KDC is assumed, to be honest, but curious and will not collude with other entities. 
%that they are no longer infected or suspected of having COVID-19.

%from infected users to be used later by the health authority in the creation of new Bloom filters containing valid keys.  The KDC considers a system initializer; thus, it does not conflict with the system's decentralization feature. The KDC is assumed, to be honest, but curious and will not collude with other entities. 

%since the KDC is just a system initializer, which is not involved in managing the parking service. \hl{Practically, the KDC can be the Ministry of Transport, for example.}

%\item 
\textbf{Consortium Blockchain.}  
Consortium Blockchain is used to run our system in a distributed manner without the need to trust a centralized entity. 
The Blockchain validators are the health authorities, and they are responsible for creating and updating two ledgers for users' COVID-19 status and zones' infection levels.

% Consortium Blockchain is used in our system to allow different health authorities to be validators and run the system despite the lack of trust between them. The Blockchain validators are responsible for creating/updating two ledgers for users' COVID-19 status and zones' infection levels.
%Specifically, the Blockchain network receives and records parking offers and reputation values on the shared ledgers
%using a pre-defined consensus algorithm.
\begin{comment}
\begin{figure}
\center\includegraphics[width=3.3in]{figures/net-mod.pdf}

\caption{The considered network model. \label{fig:Network-Model}}
\end{figure}
\end{comment}

%\item
\textbf{Health authorities.}
Health authorities in our system could be hospitals, clinics, laboratories, testing centers, etc. 
They are responsible for conducting COVID-19 tests, collecting the proof-of-contact messages collected by the infected users during the incubation period of the virus (the past 14 days), and notifying the close-contact users. 
They are also responsible for creating and updating two lists for the public keys of the infected users and the probably infected (suspected) users, and posting them on the Blockchain. 
Moreover, the health authorities need to notify the KDC with keys of infected and suspected users to stop renewing their keys until the KDC is notified by the health authorities that the users are tested negative for COVID-19.
Finally, the health authorities are responsible for categorizing zones based on their infection levels and post this information on the Blockchain.

%\textcolor{red}{Bloom filters based on the received valid keys of infected and probable infected (suspected) users with the help of KDC (i.e., health authority needs to obtain the latest keys being used by the suspected users from the KDC).} 
%Moreover, the health authorities need to inform the KDC with a list of keys of infected and suspected users to stop the renewing process of their keys and prevent them from spreading the infection elsewhere until getting proofs of negative health status from the health authorities. Furthermore, the health authorities are responsible for categorizing zones based on infection levels. The health authorities do not trust each other and perform the role of the validators in the Blockchain network. 
%The Health authority is assumed, to be honest, but curious and will not collude with other entities.

% add updating places to the things that health authority do.
%PLs publish their offers on the Blockchain. They are owned by different parties that have conflicting goals and do not trust each others. They also act as validators in the Blockchain network.
%The PLs
%can be public (e.g., commercial parking) or private where any person interested can share his residential parking slots.
%\item 
\textbf{Users.} 
The users of our system use their smartphones equipped with Bluetooth technology to exchange proof-of-contact messages when their phones are in close proximity. 
These messages are stored in the users' smartphones for the incubation period of COVID-19. 
If users get infected, they upload the proof-of-contact messages to the health authority. 
Users query the Blockchain to know whether they contacted an infected person, and also to learn the infection level of the zones before visiting them.

%Users are assumed to be honest, but curious.
%\item 

\textbf{Public places.} 
Public places in our system could be restaurants, libraries, companies, theatres, shopping centers, educational institutes, etc. 
At the entrance of these places, the infection status of each user is checked by querying the Blockchain before allowing them to access the places. 

%Places are assumed to be honest, but curious. 
%\end{itemize}

\subsection{Threat Model}
The KDC executes the system initialization and key renewal processes correctly because it can be administrated by the ministry of health which is naturally interested in controlling the virus spread. 
%Also, the KDC is considered to be independent and will not collude with any other entity in the system. 
However, the KDC is curious to learn sensitive information about the users.
The health authorities are honest-but-curious, i.e., they run the system correctly, but they are curious to learn sensitive information about the users. 
The health authorities do not trust each other and they act as validators in the Blockchain network. 
%However, they are considered to be independent and will not collude with each other or with the system users. 
Users are curious to learn sensitive information on other users, and they may also launch panic attacks by reporting false close contacts. 
In addition to the previous internal entities, adversaries can also be external attackers who try to learn sensitive information by eavesdropping on all the communications in the system.

Adversaries may try to launch the following attacks.
% The adversary intends to achieve the following goals (Note: the adversary could be one of the systems' entities or an external entity that tries to launch an attack): 

\begin{itemize}
  \item \textbf{Identification.} 
  An adversary may try to identify the real identities of the users. 
  He may also try to learn whether a person of interest is infected with COVID-19 or was in close contact of an infected user.
  
    \item \textbf{Social graph disclosure.}  
    An adversary may try to identify the social graph of an infected user, i.e., identify the close contacts of this user.
    % Health authority may try to obtain the social graph (i.e., identify contact relationships) of an infected person. 
    % The health authority may try to construct a list of close contacts of an infected user based on contact messages that have been forwarded by him to build a graph that represents the relationship and connectivity of a particular individual.  
   
    \item \textbf{Tracking.} An adversary may try to track the locations of a user using the information acquired from the messages exchanges in the system.
    
    \item \textbf{False reporting of close contacts.} An adversary may try to deceive the health authority to falsely identify users as close contacts of an infected user.
   
    \item \textbf{Message modification and manipulation.} 
    An adversary may try to modify, manipulate, or replay messages that are exchanged in the system.
    
\end{itemize}
\subsection{Design Goals}
\label{obje}
% Given the contagious nature of COVID-19 and the necessity to protect participants' privacy via thwarting various attacks while perusing infection control processes. The following is the list of goals of our proposed scheme.  
Our system is designed to achieve the following goals.

\begin{itemize}
\item \textbf{Decentralization.} Given the vulnerabilities of the centralized architectures in terms of single point of failure and DoS attacks, our system should be decentralized to enhance the availability and the reliability of the provided services. 
In other words, our system should not depend on a single central authority to conduct the contact tracing process, update the infection status of users, and categorize zones.

\item \textbf{Privacy Preservation.} 
The privacy of all users, including both infected users and their close contacts, should be preserved. 
%No one other than the health authority that has conducted the COVID-19 test should know the test result of any user. 
Specifically, no one including the health authority should identify the close contacts of any infected user.
Moreover, no one should be able to track the visited locations of any user in the system. 
Therefore, our system should be secure against privacy attacks including identification, social graph disclosure, and tracking.

% \item \textbf{Authentication.} The system should only allow legitimate users (i.e., users with a valid certificate and -ive status) to participate in the system by meeting other users and accessing places.
%anonymously without
%revealing their real identities.
\item \textbf{Infection Control.} 
Our system should control the infection spread between people as follows.
\begin{itemize}
\item \textit{Tracing and notification}. 
Our system should notify the users who closely contacted an infected user during the incubation period of the virus to quarantine.
\item \textit{Countering non-compliance}. 
Infected users and their close contacts should quarantine until they are tested negative to avoid spreading the virus.  
Our system should prevent non-compliant users (i.e., infected users and their close contacts who do not quarantine) from accessing public places and engaging with healthy people.
\item \textit{Recommending safe places}. 
Our system should classify zones based on their infection level so that users avoid contaminated zones, i.e., zones with high number of infections, to avoid spreading the virus. 
\end{itemize}

% The system should prevent risky behaviors of non-compliant users (i.e., infected and suspected users who do not act based on the received notifications of exposure and continue engaging with others rather than taking COVID-19 test) from spreading the virus. Therefore, the system should provide a digital pass based on the current COVID-19 status to be used as a tool by users in order to meet other users or access places.
% suspected users who do not act based on the received notifications of exposure and continue engaging with others rather than taking COVID-19 test.
%infected and suspicious users who do not isolate themselves and meet other users
%The system should prevent risky behaviors of non-compliant users (i.e., infected and suspicious users who do not follow the quarantine instructions) from spreading the virus by providing a digital pass based on the current COVID-19 status.

\item \textbf{Prevention of panic attacks.} 
The system should be secure against panic attacks launched by reporting false close contacts. 
To do that, the system should ensure that there were real contacts between the users and their close contacts.

% and not allow adversaries  to deceive the health authority falsely believe that an individual is one of the close contacts of someone infected.
% \item \textbf{Infectiqon integrity.}  
% The system should guarantee infection integrity (i.e., not allowing individuals to falsely claim that they are infected while they are not).
% \item \textbf{Recommendation system.} The system should recommend for users the locations with a low rate of infection to avoid contagious areas.
\item \textbf{Transparency.} 
The system should be transparent and verifiable in the processes of updating users' status and zones' infection levels.  
%categoriazitng palcess based on teh levels of infection. The process of managing parking service should be transparent to participants. Also, the process of updating the reputation scores should beverifiable and transparent to all parking lots.

\item \textbf{Efficiency and scalability.} 
In pandemics, many people usually get infected, so our system should be scalable, i.e., efficient in terms of communication, computation, and storage overheads.

\end{itemize}

\section{Proposed Infection Control System}
\label{chap:proposed scheme} 
% \begin{figure*}[!t]
% \center \includegraphics[width=7in]{figures/NetworkModel.pdf}
% \caption{An illustration for the network model} %\textit{Initialization } phase.} 
% \label{fig:NetworkModel}
% \end{figure*}
In this section,  we present our secure and privacy-preserving Blockchain-based infection control system.

\subsection{Overview}
As demonstrated in Fig. \ref{fig:NetworkModel}, there are five phases in our system, including \textit{system initialization}, \textit{contact}, \textit{tracing}, \textit{status update/check}, and \textit{public places access control}. 

In the \textit{system initialization} phase, 
the KDC distributes cryptographic credentials to the system entities. 
%Each user also obtains short-term anonymous public/private key pairs to be used as ephemeral identifiers in the \textit{contact} phase. 
Users should contact the KDC frequently to renew their credentials.
%The group of health authorities form a consortium Blockchain to store two ledgers; a COVID-19 status ledger and zones' status ledger. 
% with the each health authority and place owner registers with KDC to acquire permanent public and private key pair. Each user registers with KDC to obtain public and private key pairs (permanent and temporary key pairs). Each public key will be associated with a valid certificate for a small period of time (e.g., one day). The user will use the permanent key pair to authenticate him/herself to the KDC and to renew his/her temporary key pair. The process of updating keys will be repeated daily by users to participate in the system.
In the \textit{contact} phase, when two users encounter each other, they authenticate and check the COVID-19 status of each other, and then exchange proof-of-contact messages containing the location and time of the contact. 
%each user validates the other user's public key certificate and ensures that the user's COVID-19 status is negative. 
In the \textit{tracing} phase, if a user is tested positive for COVID-19 by a health authority, he/she should send all the contact messages collected in the past 14 days to the health authority. %Status Update \& Tracing Phase
In the \textit{status/check update} phase, 
%one of the health authorities is elected as a leader to update the Blockchain ledgers. 
%The leader collects the close contact messages from all health authorities and form two lists; one contains the temporary public keys of infected users and the other contains the temporary public keys of the close contacts. 
the health authorities use the proof-of-contact messages provided by the infected users to update the status of the infected users and their close contacts and categorize the zones based on their infection level, and post the updates on the Blockchain. 
The Blockchain is queried to check the status of users or infection level of zones.
In the \textit{public places access control} phase, 
to control the spread of the virus, before allowing a user to enter a public place and engage with other people, the infection status of the user is checked first.
%the two lists on the Blockchain are checked to make sure that the user is not infected or was in contact with infected users.
%Users can also check the zones' status before they visit locations in these zones.  

%%%%%%%%%%%%%%%%%%%%%%%%%%%%%%%%%%%%%%%%%%%%%%%%%%%%%%%%%%%%%%%%%%%%%%%%%%%%%%%%%%%%%%%%%%%%%%%%%%%%%%%%%

\subsection{System Initialization Phase} %\label{sec:System-Initialization}}
%\begin{figure*}[t]
%\center \includegraphics[width=5.5in]{Files/figures/InitializationP.pdf}
%\caption{An illustration for the \textit{system initialization} phase.} \label{fig:initialization_phase}
%%Store, contact tracing(infection control)
%\end{figure*}
The KDC generates the bilinear pairing parameters $\left\{q, \mathbb{G}, \mathbb{G}_{T}, P, e\right\}$ and chooses a secure hash function $\mathcal{H},$ where $\mathcal{H}$: $\{0,1\}^{*} \rightarrow \mathbb{G}$, $\mathbb{G}$ is a cyclic additive group of prime order $q$ and has a generator $P$, $\mathbb{G}_{T}$ is a cyclic multiplicative group of the same prime order $q$, and $e$: $\mathbb{G} \times \mathbb{G}$ $\rightarrow$ $\mathbb{G}_{T}$. 
Then, it publishes the system parameters as $\left\{q, \mathbb{G}, \mathbb{G}_{T}, P, e, \mathcal{H}\right\}$.
% In addition, each $SM_i$ chooses a secret key $x_{i} \in \mathbb{Z}_{q}^{*}$, and computes the corresponding public key $Y_{i}=x_{i} P$. Similarly, the aggregator possesses private/public key pairs $x_{g w} / Y_{g w}$, respectively.
% In the \textit{system initialization} phase, 
%two steps are required for the authentication process.
%: verification of phone number and identity authentication. each health authority will register using governmental credentials (e.g. health facility license) and obtain permanent public and private key pair. similarly, place owner will register using official papers (e.g. business license) and obtain permanent public and private key pair.
%health authorities, place owners, and users download and access the infection control app. Each user registers using his/her true identity and mobile phone number to receive an authentication code as One Time Password (OTP) through SMS. 
%After the phone number confirmation is done, the KDC sends permanent public and private key pair ($\mathcal{PK}_{U_i}$,  $\mathcal{SK}_{U_i}$) as  authorization credentials to be used by the user for future authentication. %and delete the phone number. 
Each health authority ($H$) and public place ($P$) register with the KDC to obtain long-term certified public and private key pairs $\mathcal{PK}_{H}$/$\mathcal{SK}_{H}$ and $\mathcal{PK}_{P}$/$\mathcal{SK}_{P}$, respectively, where 
$\mathcal{SK}_{H} = x_H$, $\mathcal{SK}_{P} = x_P$, $\mathcal{PK}_{H} = x_H P$, and $\mathcal{PK}_{P} = x_P P$, and $x_H, x_P \in \mathbb{Z}_{q}^{*}$. 
%Similarly, each place owner registers using official papers (e.g., business license) and obtains permanent public and private key pair ($\mathcal{PK}_{P_i}$,  $\mathcal{SK}_{P_i}$).
%each user and place register with the KDC to receive a long term public/private key pair ($\mathcal{PK}_{U_i}$, $\mathcal{SK}_{U_i}$) that is used for authorization.
Each user ($U$) should obtain a pseudo identity and short-term public and private key pair ($T_{\mathcal{PK}_{U}}$, $T_{\mathcal{SK}_{U}}$) and certificate ($Cert_{U}$), where $T_{\mathcal{SK}_{U}} = x_U$, $T_{\mathcal{PK}_{U}} = x_U P$, and $x_U \in \mathbb{Z}_{q}^{*}$.
%to be used later as ephemeral identifiers in the \textit{contact} phase. This request is signed with the user's $\mathcal{SK}_{U_i}$ for authentication purposes. 
%The KDC responds with ($T_{\mathcal{PK}_{U_i}}$, $T_{\mathcal{SK}_{U_i}}$) and an associated short-term (e.g., one day) certificate ($Cert_{U_i}$) containing the expiration time and the KDC's signature.
% using permanent public and private keys to obtain temporary public and private key pair ($Tmp_{\mathcal{PK}_U_{_i}}$ and $Tmp_{\mathcal{SK}_U_{_i}}$) and associated certificate ($Cert_U_{_i}$) from the KDC. Temporary keys only last for a small period of time (e.g., one day), and it will be used later as ephemeral identifiers in the \textit{contact}  phase.
Before a user’s certificate expires, he/she should contact the KDC to obtain a new short-term pseudo identity, public and private key pair, and certificate. 
Each user's credential should be used for a short time and no one (except the KDC) should be able to link a user's credentials to preserve the privacy of the users. 
The KDC should not renew the credentials of the infected or suspected users until they are tested negative for COVID-19.

%If the user is not infected (i.e., his/her status is negative), the KDC will send a new temporary public and private key pair to the user. 

%If the user is infected or suspected of having COVID-19 , the KDC will ignore the user's request until providing a proof (signature over his current $T_{\mathcal{PK}_{U_i}}$ from a health authority) to indicate that the status has been changed to negative. 

%Fig \ref{fig:initialization_phase} illustrates the interaction sequence of the \textit{system initialization} phase.  
%%%%%%%%%%%%%%%%%%%%%%%%%%%%%%%%%%%%%%%%%%%%%%%%%%%%%%%%%%%%%%%%%%%%%%%%%%%%%%%%%%%%%%%%%%%%%%%%%%

\subsection{Contact Phase}\label{sec:contact-phase}
\begin{figure*}[t]
\center \includegraphics[scale=0.5]{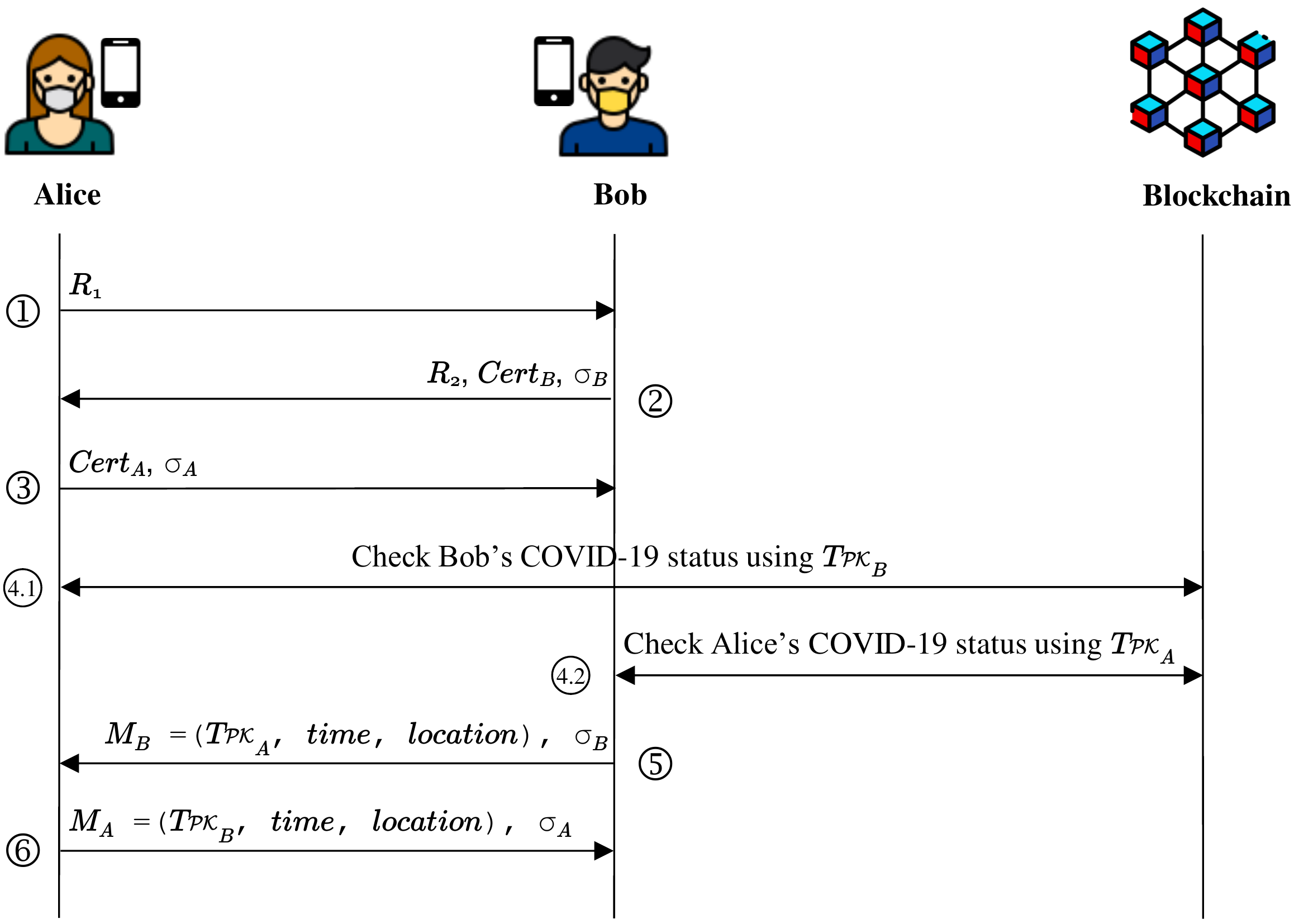}
\caption{An illustration for the exchanged messages in the \textit{contact} phase.} \label{fig:Contact-phase}
\end{figure*}
%\textcolor{red}{In this phase, each user’s smartphone is equipped with Bluetooth technology to send and receive contact messages. The contact tracing software (app) will be responsible for managing and maintaining a list of contacts, including all the messages that the user received over 14 days period.  Messages will be deleted after 14 days.} \textcolor{blue}{Here, the system will not store any personal identification information (PII) nor request for those information; thus, contact messages will not contain any sensitive data.} 
In our system, we consider both direct and indirect infections. 
Direct infection happens when an individual contracts the virus due to getting close and engaging with infected individuals, while indirect infection happens when an individual contracts the virus due to visiting public places that have been recently visited by infected individuals. 
In the latter case, individuals may get infected by touching surfaces contaminated with COVID-19 and then touching their mouths, noses, or eyes. 

\textbf{Direct infections.}
Before two users get closer to each other or engage in any activity, each user needs to make sure that the other user is healthy (non-infected). 
To do so, each user first checks that he/she is meeting a legitimate user with a valid public key certificate. 
Then, each user needs to check the COVID-19 status of the other user by sending a request to the Blockchain with the public key of the other user. 
If the Blockchain returns that the COVID-19 status of the two users are negative, 
they use Bluetooth to exchange proof-of-contact messages to be stored in their smartphones. 
These messages will be used later to notify the users that were in close contact within the last 14 days to someone who is tested positive for COVID-19. 

% \textcolor{red}{This is done by performing a lookup operation to check the Bloom filter for $Tmp_{\mathcal{PK}_U_{_i}}$ entry. If there was no record of the receiving key (other user's key) in the Bloom filter, then the user is not suspected of having/carrying COVID-19 and can pair the two devices to obtain from him/her contact message $M_U_{_i}$. However, if any of the two users failed to prove that the shared key is valid and belongs to the individual himself (i.e., the one that currently in meeting with), then the connection will be aborted by the other user.} 
In the following, we explain the steps of the \textit{contact} phase executed by two users (Alice and Bob). 
The messages exchanged between the system entities involved in this phase are shown in Fig. \ref{fig:Contact-phase}:

\textbf{Step} \raisebox{.5pt}{\textcircled{\raisebox{-.9pt} {1}}}: Alice sends a random number $R_1$ to Bob.

\textbf{Step} \raisebox{.5pt}{\textcircled{\raisebox{-.9pt} {2}}}:  \begin{itemize}
   
    \item Bob uses his temporary private key, $T_{\mathcal{SK}_B}= x_B$, to sign $R_1$ as follows.
    \begin{equation} \label{hash}
    \sigma_{B}=x_{B} \mathcal{H}(R_1).
    \end{equation}
    
 \item Bob sends the certificate of his temporary public key ($Cert_B$) and signature ($\sigma_B$) to authenticate himself to Alice. Moreover, Bob sends a random number $R_2$ to Alice to sign so that he can authenticate Alice.
%($Cert_{\mathcal{Tmp}_PK}$)($Cert_t_m_p_p_k$) $Cert_{\mathcal{T}_PK}$ &&&&&&&&& $\mathrm{\sigma}$  \[ \sigma  \]  $\mathrm{\sigma}$
   
\item Alice validates the certificate of Bob ($Cert_B$) by verifying the KDC's signature on the certificate using the KDC's public key, and checking whether the certificate is expired.
    % \item Check the start and expiration dates of the certificate whether configured properly (e.g., the certificate should be only valid for one day).
%However, if the certificate is expired, this may be because the user did not renew the certificate and this may be because the KDC denied   
%indicates that the user is a malicious user or suspended user (i.e., a user that is not allowed to participate in the system with COVID-19 status +ive or suspected).

\item If the certificate is valid, then Alice proceeds by verifying Bob's signature; otherwise, Alice should not continue the protocol or engage with Bob in any activity. 
Alice uses the public key included in Bob’s digital certificate, $T_{\mathcal{PK}_B}=Y_B$, to verify the received signature ($\sigma_B$) by checking the following.
\begin{equation} \label{VER}
    e\left(\sigma_{B}, P\right) \stackrel{?}{=} e\left(\mathcal{H}(R_1), Y_B\right)
\end{equation}
%%%%%%%%%%%%%%%%%%%%%%%%%%%%%%%%%%%%%%%%%%%%%%%%%%%%%%%%
% explin the verify method ( how it work)   \\ H(R) \stackrel{?}{=}
%%%%%%%%%%%%%%%%%%%%%%%%%%%%%%%%%%%%%%%%%%%%%%%%%%%%%%%%
\textbf{Proof.}
	\begin{align*}
	e(\sigma_{B}, P) &\stackrel{}{=} e(x_{B} \mathcal{H}(R_1), P)\\
	&\stackrel{}{=} e(\mathcal{H}(R_1), x_{B} P)\\
	&\stackrel{}{=}e\left((\mathcal{H}(R_1), Y_{B}\right).
   \end{align*}
% (ownership verification and $\mathcal{PK}$%PK 
%  validation process). 
 
   \end{itemize}
   
\textbf{Step} \raisebox{.5pt}{\textcircled{\raisebox{-.9pt} {3}}}: Alice responds to Bob with the certificate of her temporary public key ($Cert_A$) and signature ($\sigma_A$) on both $R_1$ and $R_2$.
Similarly, Bob verifies $Cert_A$ and $\sigma_A$.

 \textbf{Step} \raisebox{.5pt}{\textcircled{\raisebox{-.9pt} {4}}}: Alice checks the COVID-19 status of Bob by sending a request to the Blockchain containing Bob’s temporary public key $T_{\mathcal{PK}_B}$. 
The Blockchain returns ``infected'', ``close contact", or ``not found" based on the status of the user.
If the status of Bob is ``infected" or ``close contact", then Alice terminates the protocol and should not get closer to Bob or engage in an activity with him. 
However, if the status is ``not found", this indicates that it is safe for Alice to meet Bob. 
Similarly, Bob repeats these steps to check the COVID-19 status of Alice.
%In case that the COVID-19 status of both Alice and Bob are ``not found", they complete the protocol by exchanging proof-of-contact messages. 
%On the other hand, if $T_{\mathcal{PK}_B}$ is found in any filter, Alice will abort the protocol.
% Because it is an indication that Bob’s COVID-19 status is negative, and it is safe to meet him. Otherwise, if Bob’s temporary public key was found in any filter, Alice will abort the protocol.

\textbf{Step} \raisebox{.5pt}{\textcircled{\raisebox{-.9pt} {5}}}:\begin{itemize}
   \item Bob composes a proof-of-contact message ($M_B= T_{\mathcal{PK}_A}, time, location$), containing the public key of Alice and the time and location of the contact.  
   Then, Bob sends $M_B$ and a signature on it ($\sigma_B= x_B \mathcal{H}(M_B)$) to Alice.
% \begin{multline}
% M_B=T_{\mathcal{PK}_A}, time, location,x_B \mathcal{H}(time || location || T_{\mathcal{PK}_A}), 
% \end{multline}
% The parameters $time$ and $location$ indicate the time and location of the contact of Alice and Bob, and $x_B \mathcal{H}(time || location || T_{\mathcal{PK}_A})$ is Bob's signature on Alice's temporary public key $T_{\mathcal{PK}_A}$, time and location. This signature is needed as a proof for the contact of Alice and Bob to prevent the false reporting (or panic) attack and also to guarantee the authenticity and integrity of the received message.
The signature $\sigma_B$ is needed as a proof for the contact of Alice and Bob to prevent the false reporting (or panic) attack and also to guarantee the authenticity and integrity of the message.
%The signature over Alice's key, time and location is for the prevention of false reporting and the confirmation of the received message's authenticity and integrity.
\item Alice accepts $M_B$ if the signature $\sigma_B$ is valid. 
Then, the message $M_B$ associated with the signature $\sigma_B$ and Bob's temporary public key certificate ($M_B, \sigma_B, Cert_B$) is stored in Alice's smartphone.
 \end{itemize}
   
% \item Alice accepts $M_B$ if the received signature $\sigma_B$ is valid. 
% Then, the message $M_B$ associated with the signature $\sigma_B$ and Bob's temporary public key ($M_B, \sigma_B, T_{\mathcal{PK}_B}$) is stored in Alice's smartphone.
%\textcolor{red}{and  Bob will execute the protocol in a similar manner.}
%and $Tmp_{\mathcal{PK}_A}$ is Alice's temporary public key. Bob
% \item Similarly, Alice sends a contact message $M_A$ and a signature $\sigma_A$ to Bob. If the signature $\sigma_A$ is valid, the tuple ($M_A, \sigma_A, T_{\mathcal{PK}_A}$) is stored in Bob's smartphone.
\textbf{Step} \raisebox{.5pt}{\textcircled{\raisebox{-.9pt} {6}}}: Alice, similarly, sends a contact message $M_A$ and a signature $\sigma_A$ to Bob. If the signature $\sigma_A$ is valid, the tuple ($M_A, \sigma_A, Cert_A$) is stored in Bob's smartphone.
\vspace{2mm}

% Note that the aforementioned steps (1 to 8) are executed simultaneously by the two users (Alice and Bob). Therefore, in step 7, Bob already has Alice's temporary public key to include it in the contact message. The steps included in the \textit{contact} phase are illustrated in Fig.~\ref{fig:Contact-phase}. 
% Moreover, in our system, we consider the indirect infection that may happen by visiting a place that has been just visited by an infected person.
\textbf{Indirect infections.}
When a user visits a public place, regardless of encountering other users, he/she should exchange a proof-of-visiting message with the place itself, which indicates that he/she has visited the place in a specific time period. 
When Alice visits a restaurant $R$, she receives a proof-of-visiting massage ($M_R=T_{\mathcal{PK}_A}, time, location$) and a signature on it ($\sigma_R= x_R \mathcal{H}(M_R)$).
% The tuple ($M_R, \sigma_R, \mathcal{PK}_R$) is stored in Alice's smartphone if the received signature $\sigma_R$ is valid, where $\mathcal{PK}_R$ is the public key of the restaurant. Similarly, a proof-of-visiting tuple ($M_A, \sigma_A, \mathcal{PK}_A$) is stored in the restaurant's local storage.
The tuple ($M_R, \sigma_R, Cert_R$) is stored in Alice's smartphone if the received signature $\sigma_R$ is valid, where $Cert_R$ is the public key certificate of the restaurant. Similarly, a proof-of-visiting tuple ($M_A, \sigma_A, Cert_A$) is stored in the restaurant's local storage.

In order to reduce the storage overhead on the users' smartphones, we use an aggregate signature technique. 
Every time a user receives a signed message either from other users or visited places, he aggregates the signature of this message with the signatures stored in his smartphone. 
Assume that the user $U$ has collected $\mathcal{M}$ messages. Instead of storing $\mathcal{M}$ individual signatures, the user stores only one aggregate signature ($\sigma_{agg}$), where the size of $\sigma_{agg}$ is similar to the size of each individual signature. $\sigma_{agg}$ is calculated as follows.
% For the sake of efficiency, the infected user (U) aggregates all signatures in the received messages from close contacts and visited places into one signature. Assume that the number of messages is $\mathcal{M}$, then the aggregate signature ($\sigma_{agg}$) is calculated as follows. 
\begin{equation}
\label{aggregate}
    \sigma_{agg}=\sum_{i=1}^{\mathcal{M}} x_i \mathcal{H}(T_{\mathcal{PK}_{U_i}}, time_i, location_i)
\end{equation}
%Note that the aforementioned steps (steps 1 to step 7) are executed simultaneously by the two users (Alice and Bob).
% \begin{figure*}[t]
% \center \includegraphics[width=5.5in]{figures/ContactP.pdf}
% \caption{An illustration for the \textit{Contact} phase.} \label{fig:Contact-phase}
% \end{figure*}
% \begin{figure}
% \center \includegraphics[width=3in]{Files/figures/ContactPP.pdf}

% \caption{An illustration for the \textit{contact} phase: visited places.} \label{fig:visiting_places}

% \end{figure}
%%%%%%%%%%%%%%%%%%%%%%%%%%%%%%%%%%%%%%%%%%%%%%%%%%%%%%%%%%%%%%%%%%%%%%%%%%%%%%%%%%%%%%%%%%%%%%%%%%%%%%%%%%

\subsection{Tracing Phase} \label{sec:Reporting_Phase}
%\begin{figure}
%\center \includegraphics[width=\columnwidth]{Files/figures/TRACING2.pdf}
%\caption{An illustration for the exchanged messages in the \textit{tracing} phase.} \label{fig:Tracing_phase}
%\end{figure}

In this phase, the health authority collects all the proof-of-contact and proof-of-contact messages from the infected users. 

\textbf{Direct infections.} 
If a user is tested positive for COVID-19, he/she should send the proof-of-contact and proof-of-visiting messages collected in the last 14 days to the health authority to notify the close-contact users. 
Assume that the number of collected messages by the user in the last 14 days is $\mathcal{N}$. 
These messages are stored in the user's smartphones as groups of $\mathcal{M}$ messages, where each group is associated with an aggregate signature $\sigma_{agg}$. 
When a user sends his messages to the health authority, he sends them at different times (only one group at a time) so that the health authority cannot know if the messages are sent from same user or different users because it cannot link the short-term public keys used by the same user.
% Thus, in 

%\textbf{Step} \raisebox{.5pt}{\textcircled{\raisebox{-.9pt} {5}}}: 

The user $U$ uploads the following data to the health authority: $\{M_1, M_2, \dots, M_\mathcal{M}\}$ and $\sigma_{agg}$, where $M_i=(T_{\mathcal{PK}_{U_i}}, time_i, location_i, Cert_i)$. 
Using the aggregate signature reduces the communication overhead because the user uploads only one signature instead of uploading $\mathcal{M}$ individual signatures.
Once the user uploads proof-of-contact messages, the health authority needs to check the authenticity and integrity of all messages to make sure that the user is not launching a false reporting attack. 

%The temporary public key certificates included in the messages protects the system against the malicious users that may try to upload forged messages with keys of non-existing users to consume the system resources.
% where $M_i=(T_{\mathcal{PK}_{U_i}}, time_i, location_i, Y_i)$.
% Instead of uploading $\mathcal{M}$ individual signatures to the health to verify them, the infected user can only uploads $\sigma_{agg}$. Thus, the data uploaded by the infected user to the health authority are $\sigma_{agg}$ and $\{M_1, M_2, \dots, M_\mathcal{M}\}$, where $M_i=(Y_i, time_i, location_i, T_{\mathcal{PK}_{U_i}})$.
Moreover, using the aggregate signature reduces the computation overhead because the health authority verifies only one signature $\sigma_{agg}$ instead of verifying $\mathcal{M}$ signatures. If $\sigma_{agg}$ is valid, this indicates that all the underlying individual signatures are valid. To verify the aggregate signature $\sigma_{agg}$, the health authority checks the following.
\begin{equation}
e\left(\sigma_{agg}, P\right) \stackrel{?}{=} \prod_{i=1}^{\mathcal{M}} e\left(\mathcal{H}(T_{\mathcal{PK}_{U_i}} || time_i || location_i), Y_{i}\right)
\end{equation}
\textbf{Proof.} \\
	\begin{align*}
	e\left(\sigma_{agg}, P\right)
	&\stackrel{}{=} e\left(\sum_{i=1}^{\mathcal{M}} x_i \mathcal{H}(T_{\mathcal{PK}_{U_i}}, time_i, location_i), P\right)\\ 
	&\stackrel{}{=} \prod_{i=1}^{\mathcal{M}} e\left(x_i \mathcal{H}(T_{\mathcal{PK}_{U_i}}, time_i, location_i), P\right)\\
&\stackrel{}{=} \prod_{i=1}^{\mathcal{M}} e\left(\mathcal{H}(T_{\mathcal{PK}_{U_i}}, time_i, location_i), x_i P\right)\\
	&\stackrel{}{=}\prod_{i=1}^{\mathcal{M}} e\left(\mathcal{H}(T_{\mathcal{PK}_{U_i}}, time_i, location_i), Y_{i}\right).
   \end{align*}
% The temporary public keys of the users included in the collected messages from the infected user will be used in the \textit{status update phase}. This is to update the COVID-19 status of those users as probable infected due to coming into contact with an infected user.

% In case that the received $\sigma (Y)$ is valid and the public key certificate ($Cert_B$) has a valid KDC signature but it is expired, the health authority searches for Bob's temporary public key $T_{\mathcal{PK}_{B}}$ in the two Bloom filters. If it is found in any filter, this means that the latest COVID-19 status of Bob is either positive or suspected. In this case, Bob is tested and if the result is negative, he obtains a signature from the health authority on his $T_{\mathcal{PK}_{B}}$ as a proof of recovery and $T_{\mathcal{PK}_{B}}$ should be removed from the Bloom filter. If $T_{\mathcal{PK}_{B}}$ is found in the filter of the suspected users and the test result is positive, 
% % \begin{figure}
% \center \includegraphics[width=3.8in]{figures/ReportingP.pdf}
% \caption{An illustration for the \textit{Reporting } phase.} \label{fig:Reproting_phase}
% \end{figure}

\textbf{Indirect infections.}
After verifying the proof-of-visiting messages containing the places visited by the infected users in the last 14 days, the health authority traces the probable indirect infections. 
To do that, it requests these places to send the proof-of-visiting messages collected from the users that visited the places at the same time frame as the infected user. 
Also, because COVID-19 virus can survive on surfaces for several hours, the place can also send the proof-of-visiting messages for several hours after the visit of the infected user.
%These users are probably infected because of their existence in the same place and time with an 
%as messages are used by the health authority to identify the probable infected users due visiting contaminated places. 

The collected information in this phase is used in the \textit{status update/check phase} to update the COVID-19 status of the infected users and their close contacts, and also categorize the infection level of the different zones.

%%%%%%%%%%%%%%%%%%%%%%%%%%%%%%%%%%%%%%%%%%%%%%%%%%%%%%%%%%%%%%%%%%%%%%%%%%%%%%%%%%%%%%%%%%%%%%%%%%%%%%%%%%

\subsection{Status Update/Check Phase}
\begin{figure*}[!t]
\center \includegraphics[scale=0.5]{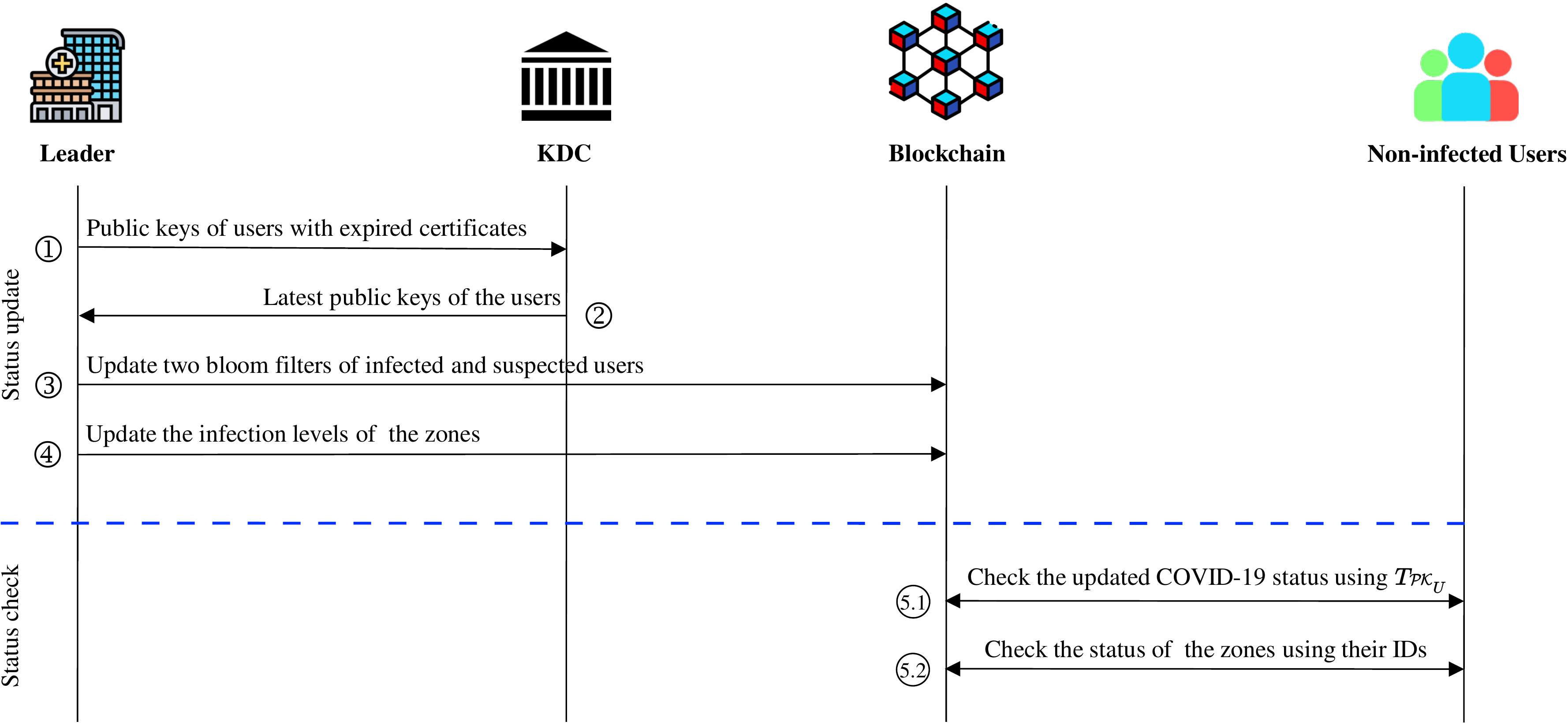} 
\caption{An illustration for the \textit{status update/check} phase. \vspace{-2mm}} \label{fig:StatusUpdateP}
\end{figure*}

\subsubsection{Status Update} 
After a health authority verifies the received messages from infected users and their visited places, it forwards these messages to the other health authorities which act as validators in the Blockchain network to verify them.
One of these health authorities is elected to act as a leader based on the Raft consensus algorithm. 
The leader adds the public keys of the infected and close-contact users in two Bloom filter lists; one contains the keys of the infected users and the other contains the keys of the close contacts users.
% , and sends them to the KDC without transmitting the actual messages and test results.  %\textcolor{green}{$Tmp_{\mathcal{PK}_U_{_i}}$ of infected users , where $ (1\leq	i\leq	N) $, $N$  is the total number of infected users. And their contact messages ($M_{\mathcal{C}_U_{_i}}= M_U_{_1},M_U_{_2},M_U_{_3},...M_U_{_X}$), where $ (1 \leq i \leq X) $,  $X$ is total number of close contacts messages a $U_i$  received within 14 days period, visited places' messages  ($M_{\mathcal{P}_U_{_i}}= M_P_{_1},M_P_{_2},M_P_{_3},...M_P_{_J}$),where $ (1\leq	i\leq	J) $, $J$ is the total number of messages $U_i$ received from places in 2 weeks duration, to the other health authorities in the Blockchain network to verify the messages. Based on the selection of a miner (new leader) to generate a block (see \ref{sec:Blockchain} and block generation section) and communicate with KDC.} 
% Because a contact may happen in the 14 days and due to using short-term certificates, some public keys may have been expired and the user renewed it. In this case, to obtain the latest public key of the user, the health authority retrieves the latest public key from the KDC without revealing the location and time of the contact and also the public key of the other user in the contact. This is important to preserve privacy be because the KDC can link the short-term keys to the real identity of the users and by receiving these information it can know who met whom, where, and when.     
Because a contact may have happened in the past 14 days and due to using short-term certificates, some public keys may have expired and the users have renewed them. 
In this case, to obtain the latest public key of a user, the leader retrieves it from the KDC without revealing the location and time of the contact and also the public key of the other user in the contact. 
This is important to preserve privacy because the KDC can link the short-term keys to the real identity of the users and by receiving these information it can know who met whom, where, and when and build a social graph.   
%The leader groups the public keys of the users and all contact messages into two lists; one contains the keys of the infected users and the other contains the keys of the suspected users, and sends them to the KDC without transmitting the actual messages and test results. 

Fig. \ref{fig:StatusUpdateP} illustrates the steps of updating the  users' COVID-19 status and the infection level of the zones. 
These steps are explained as follow.

\textbf{Step} \raisebox{.5pt}{\textcircled{\raisebox{-.9pt} {1}}}: The leader sends the public keys of the users with expired certificates to the KDC.

\textbf{Step} \raisebox{.5pt}{\textcircled{\raisebox{-.9pt} {2}}}: The KDC stops renewing the keys of those users until they are tested negative for COVID-19, and responds to the leader with the latest public keys of the users. 

% the lists from the leader, it stops the key renewal process for the users whose keys are included in the lists. Afterward, the KDC responds to the leader with updated lists containing the latest temporary public keys of these users.
% an updated list contains the latest keys of the suspected users (i.e., replaces old keys (invalid keys) with new ones (valid keys)). 
\textbf{Step} \raisebox{.5pt}{\textcircled{\raisebox{-.9pt} {3}}}:
% The leader broadcasts the updated lists to the other health authorities in the Blockchain network (followers).
% \textcolor{blue}{A copy of the temporary public keys of infected and suspected are recorded locally by health authorities.}
% Next, 
The leader updates two Bloom filters using the latest keys; one for the infected users and the other for the suspected users. 
% the leader creates two Bloom filters, one contains the keys of the infected users and the other contains the latest keys of the suspected users as shown in Fig. \ref{}.
These Bloom filters are broadcasted to the follower validators in the form of a proposed block to be appended to the COVID-19 status ledger.
% The followers can verify the new  block by ensuring that the keys of the infected and suspected users are contained in the corresponding Bloom filters. 
Each follower votes for the new block and the leader waits for the majority of the votes to add the new block to the ledger. 
Finally, the new block is signed by the leader and broadcasted to the followers to be added to their copies of the ledger. 
%The integrity of the block is achieved through the the leader’s signature as it prevents the attackers from manipulating the blocks because they do not have the leader's private key.

% Suppose the test result of a suspected user turns/has been changed to negative. In that case, the user will need to obtain proof from the health authority to renew his/her key and continue participating in the system. The health authority should already have the user’s $Tmp_{\mathcal{PK}_U_{_i}}$ as all the health authorities keep the latest list of temporary public keys of all infected and suspected users. Therefore, when a suspected/infected user wants to take/retake COVID-19 test, he/she will send his/her $Cert_U_{_i}$. The health authority will acquire $Tmp_{\mathcal{PK}_U_{_i}}$ from the $Cert_U_{_i}$ and search the local temporary public keys' list (i.e., search the list of +ive and suspected keys) after the confirmation of $Tmp_{\mathcal{PK}_U_{_i}}$ in the list. Then the health authority will perform the test, assuming that the test result turns negative. The health authority will sign over the $Tmp_{\mathcal{PK}_U_{_i}}$ as an indication of changing a user’s status to negative. Due to the health authority signature, the KDC will have to renew the user’s key pair to precede his/her daily life activities.
\textbf{Step} \raisebox{.5pt}{\textcircled{\raisebox{-.9pt} {4}}}: In our system, each city is divided into small geographic areas called zones, where each zone has a unique identifier. 
%These zones are categorized to red, orange, and green based on the number of confirmed and probable infections discovered in these zones. 
To determine the category of a zone, the weighted average of the number of infected and suspected users in the last 14 days is calculated and compared to two thresholds ($Th_1$ and $Th_2$), where $Th_2 > Th_1$. 
If the weighted average of a zone is greater than $Th_2$, then the zone's status is \textit{red}. 
If the weighted average of a zone is greater than $Th_1$ and less than $Th_2$, then the zone's status is \textit{orange}. 
Otherwise, the zone's status is \textit{green}.
%The leader is the entity that is responsible also for updating the status of the zones. 
% categorizing zones based on the infection levels occurred in the zones calculated from the numbers of infected and suspected users discovered in these zones.
After that, the leader creates a new block containing the zones' identifiers (IDs), their updated average infection numbers, and their updated status. 
This new block is broadcasted to the Blockchain followers to be approved and appended to the zones status ledger.
%  The followers can verify the block's content as they have proof-of-contact and proof-of-visiting messages.
% The steps included in the \textit{status update} phase are illustrated in Fig. \ref{fig:StatusUpdateP}.

% Bloom filters are used because they store data compactly. 
% The first Bloom filter will contain valid keys of infected users with positive COVID-19 status. The second Bloom filter will contain valid keys of probable infected users with suspected COVID-19 status.}
\begin{figure*}[!t]
\center \includegraphics[scale=0.5]{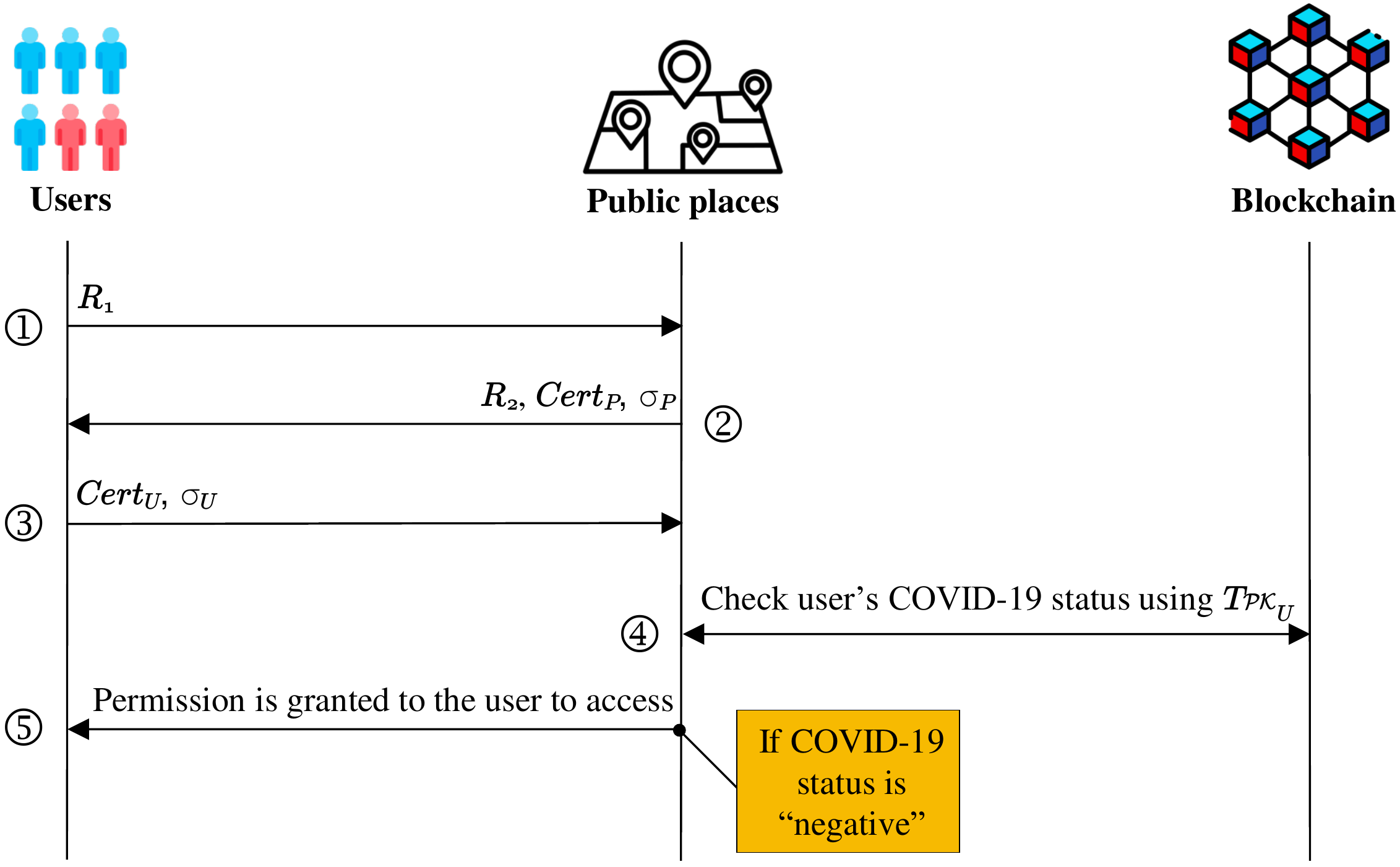}
\caption{An illustration for the \textit{public places access control} phase.\vspace{-3mm}} \label{fig:COVID-19StatusVerification_phase}
\end{figure*}

\subsubsection{Status check} 
Fig. \ref{fig:StatusUpdateP} shows the exchanged messages between the users and the Blockchain to check the COVID-19 status of the users and zones status.

\textbf{Step} \raisebox{.5pt}{\textcircled{\raisebox{-.9pt} {5}}}: 
A non-infected user $U$ can check his/her updated COVID-19 status by sending a request to the Blockchain containing his/her latest temporary public key $T_{\mathcal{PK}_U}$. 
If $T_{\mathcal{PK}_U}$ is found in the Bloom filter of suspected users, this indicates that the user has closely contacted (in the last 14 days ) someone who is tested positive for COVID-19, and thus he/she needs to quarantine immediately and visit a health authority to take COVID-19 test.
% When a suspected/infected user goes to take/retake COVID-19 test, the health authority sends a random number $R$ as a challenge to the user. The user responds with a signature ($\sigma (Y)$), calculated according to Eqs. \ref{hash}-\ref{sig}, and his temporary public key certificate ($Cert_{U_i}$). In case that the received $Cert_{U_i}$ is invalid (expired) and the received $\sigma (Y)$ is valid , the health authority searches for the user's temporary public key $T_{\mathcal{PK}_{U_i}}$ in the two local lists.
% % \textcolor{blue}{the health authorities should search the local list as the certificate is expired, and invalid keys should not be kept in the Bloom filter}.
% If it is found in any list, this means that the latest COVID-19 status of the user is either suspected or positive. In this case, the user is tested and if the result of the test is negative, the user obtains a signature from the health authority on his $T_{\mathcal{PK}_{U_i}}$ as a proof of recovery.
If a suspected user is tested negative, the health authority should make a transaction to remove the temporary public key of this user from the list of suspected users (close contacts). 
%Also, the user obtains a signature from the health authority on his temporary public key as a proof of recovery.
% and $T_{\mathcal{PK}_{U_i}}$ should be removed from the Bloom filter \textcolor{blue}{Bloom filter only contain valid keys, expired keys are not included in the new filter
% }.
Once a user is removed from the lists of infected or suspected users, the KDC resumes renewing his/her keys so that he/she can access public places. 
%After that, the user can provide this proof to the KDC to issue a new public and private key pair for the user to proceed his/her daily life activities. 
On the other hand, if the suspected user is tested positive, the user is required to send his/her messages collected in the last 14 days, and his/her temporary public key should be moved to the Bloom filter of infected users. 
Furthermore, non-infected users can check the status of a zone they want to visit by sending a request to the Blockchain containing the zone's ID. 

\subsection{Public Places Access Control Phase} \label{Cov-phase}

In this phase, a public place checks COVID-19 status of the users before allowing them to access the place. 
If the status is negative, the user can enter the place; otherwise, the user is not allowed to enter the place to prevent spreading the virus. 
Therefore, the COVID-19 status is used as a \textit{digital pass} to prevent infected and suspected users from accessing public places. 
Fig. \ref{fig:COVID-19StatusVerification_phase} shows the messages exchanged between users and public places. 
Specifically, when a user visits a public place, the following steps are executed.
% If the status is +ive or suspected, whether with a valid or invalid digital certificate, the user cannot enter the place.
% Therefore, the status is a digital pass tool and can be used between users in the \textit{contact} phase to check their current health status. The following are the steps of COVID-19 status verification process:
%Therefore, the status is used as a digital pass. This tool is used between users in the contact phase to check their current health status. The following are the steps of COVID-19 status verification process:

\textbf{Step} \raisebox{.5pt}{\textcircled{\raisebox{-.9pt} {1}}}: The user sends a random number ($R_1$) to the public place. 

\textbf{Step} \raisebox{.5pt}{\textcircled{\raisebox{-.9pt} {2}}}: The place signs $R_1$ and sends its certificate $Cert_P$, the signature $\sigma_P$, and a random number ($R_2$) to the user. Then, the user verifies the certificate $Cert_{P}$ and the $\sigma_P$. 
        % \item The user checks the validity of the received certificate $Cert_{P}$ and if it is invalid, the user is not allowed to enter the place. If $Cert_{U}$ is valid, the place verifies the received $\sigma_U$ according to Eq. \ref{VER}.

\textbf{Step} \raisebox{.5pt}{\textcircled{\raisebox{-.9pt} {3}}}: The user signs $R_2$ and sends his/her temporary public key certificate $Cert_U$ and the signature $\sigma_U$ to be authenticated by the place.

% validation of the transmitted certificate $Cert_U_i$ (steps of certificate validation mentioned in \ref{sec:contact-phase}).
% \item Place owner will verify the signature as follow:
% \begin{equation} 
%   \begin{aligned}
%      \begin{split}
%       Accept/Reject \leftarrow Verify(Tmp_{\mathcal{PK}_U_{_i}}, \sigma,Y) \\
%     % \textcolor{blue}{ Verify(Tmp_{\mathcal{PK}_B}, \sigma,Y) =Dec_{\mathcal{TMP}_{\mathcal{PK}_B} }(Y)}\\
%     %\textcolor{blue} {Dec_{\mathcal{TMP}_{\mathcal{PK}_B} }(Y) \stackrel{?}{=} H(R)}
%      \end{split}
%      \end{aligned}
%      \end{equation}
%      If Accept, then the inputs $\sigma$ and $Y$ are valid.
\textbf{Step} \raisebox{.5pt}{\textcircled{\raisebox{-.9pt} {4}}}: After the mutual authentication, the place checks the COVID-19 status of the user by sending a request to the Blockchain containing the user’s temporary public key $T_{\mathcal{PK}_U}$. 
     
\textbf{Step} \raisebox{.5pt}{\textcircled{\raisebox{-.9pt} {5}}}: If $T_{\mathcal{PK}_{U}}$ is not found in any of the two filters, this indicates that the COVID-19 status of the user is negative, and thus the user can access the public place. 
Otherwise, the user is banned from accessing the place to avoid spreading the virus to other users.

%In this phase, the place owner/manager can check COVID-19 status of users (customers/ workers) in-order to allow them to access the place. If the status is -ive, the user will not have a record in the Blockchain, and he/she can access the place. If the status is +ive or suspected, the user will have an entry in the Bloom filter in Blockchain as long as his/her certificate is valid. When the certificate becomes invalid, the infected/suspected temp will no longer be in the Bloom filter, and the user cannot enter places due to posses of an invalid certificate. Therefore, status is used as a digital pass. This tool is used between users in the contact phase to check their current health status. The following are the steps of COVID-19 status verification process:

%\section{Performance Evaluation}

%\begin{comment}
%In this section, we evaluate our proposed scheme. We first analyze the security and privacy features of our scheme, followed by communication and computation overheads.

\section{Security and Privacy Analysis}
\label{chap:securityandprivacy}

In this section, we analyze the security and privacy features provided by our system.

% \textit{\textbf{Preposition 1.}   The proposed system can manage the update of COVID-19 status  and categorize zones based on infection levels without relying on a single central authority.
% }% decentralization
\vspace{2mm}\textit{\textbf{Preposition 1.} Our system ensures the availability and reliability of the provided services.
}% decentralization

% \textbf{Proof.} The system consists of multiple subsystems with various functionaries that can be managed without counting on a single entity or third party to provide these services. This is accomplished by utilizing Blockchain technology to record temporary public keys of infected and suspected users and categorize zones according to infection levels. This makes the system more resilient in facing issues of centralized systems, including DoS attack and single point of failure problem.
\textbf{Proof.} Instead of depending on a single central authority to update the COVID-19 status of the users and categorize zones based on infection levels, our system  is decentralized and run by a group of health authorities that form a consortium Blockchain. 
Utilizing the Blockchain technology makes our system resilient against the single point of failure problem and the DoS attacks threatening the availability of the provided services. 
This is because for an adversary to make the services unavailable, it is not sufficient to attack a single node as it is the case in the centralized system. 
Moreover, utilizing the Blockchain enhances the reliability of the provided services. 
This is because it is impossible for an adversary to change the content of the Blockchain ledger without compromising the majority of the Blockchain nodes.

\vspace{2mm} \textit{\textbf{Preposition 2.} Our system thwarts the identification attack that aims to disclose a user’s COVID-19 status.
}%identification attack

\textbf{Proof.} In our system, the real identities of the users are hidden. 
Instead, unlikable pseudo-identities and short-term public keys are used to provide anonymity to the users.
In the COVID-19 \textit{status update/check} phase of our system, only the latest keys of the infected and suspected users are stored in the Bloom filters.
Therefore, if an attacker wants to track the COVID-19 status of a particular user, he cannot do that because he does not know the short-term public keys used by the user. However, for traceability of COVID-19 infections, when two users encounter each other, they need to exchange their short-term public keys. 
%Thus, it is not possible for any health authority or individual to track the COVID-19 status for any user in the system because it is associated with his/her temporary public keys that change frequently. 
%No entity except the KDC can link the temporary public key of a certain user or link these keys to his true identity. 
Moreover, if a user's status is updated to ``suspected", he cannot learn the user who infected him because users frequently change their keys and only the last key (which is not linkable to the old keys) is stored in the Bloom filter on the Blockchain.

% Suppose a user ($U_i$) has received a certificate ($Cert1_{U_j}$) from a user ($U_j$) that contains $T_{\mathcal{PK}_{U_j}}$. If the two users encounter each other again in the future, $U_i$ will receive chances that $U_i$ will receive the same certificate containing the same $Tmp_{\mathcal{PK}_U_{_j}}$ of $U_j$  are very low as the key probably has been changed at the time of the second encounter. 
% The KDC is the only entity that can link the user’s credentials (e.g., permanent and temporary keys) to the user’s true identity. Future temporary keys of $U_j$  are unknown to $U_i$, and Bloom filters only contain the latest keys of infected and suspected users; thus, $U_i$  can not track or identify $U_j$  COVID-19 status. 

\vspace{2mm}\textit{\textbf{Preposition 3.} 
Our system thwarts the social graph disclosure and tracking attacks. 
}% social graph

\textbf{Proof.} 
The social graph disclosure attacks aim to identify the close contacts of an infected user, and the tracking attacks aim to track users’ movements and activities using the location data of the proof-of-contact and proof-of-location messages.
In the \textit{tracing} phase of our system, if a user is tested positive for COVID-19, he sends the proof-of-contact and proof-of-visiting messages collected from his/her close contacts and visited places to the health authority. 
Using these messages, the healthy authority cannot create a social graph for the close contacts of the user.  
This is because pseudo-identities and short-term public keys are used, so if the user meets another user multiple times, the health authority cannot figure out this information because it cannot link the short-term public keys used by the same user.    
Moreover, when a user sends his proof-of-contact and proof-of-visiting messages to the health authority, he sends them at different times so that the health authority cannot know if the messages are sent from same user or different users to prevent it from even learning the pseudo-identities of a user's close contacts.    
Moreover, since the KDC can link the short-term public keys of each user, when the health authority resolves the latest public keys of the users in the \textit{status update/check} phase, it only forwards the public keys to the KDC and it does not forward the location information to prevent the KDC from learning the locations visited by each user. 
The health authority should also shuffle the public keys of different users before sending them to the KDC so that it cannot learn the users who contacted each other.

\vspace{2mm}\textit{\textbf{Preposition 4.} Our system allows only authorized and healthy (non-infected) users to interact with other users and access public places.}%Authentication

\textbf{Proof.} 
The KDC issues a certified public and private key pair ($T_{\mathcal{PK}_{U_i}}$, $T_{\mathcal{SK}_{U_i}}$) for each user to be able to participate in the system, i.e., be able to interact with other users and access public places. 
Each user needs to authenticate to other users to interact with them and authenticate to the public places to access them.
Therefore, if a user does not have valid credentials, he cannot participate in the system.  
Moreover, if a user is tested positive for COVID-19 or closely contacted an infected user, he cannot contact other users or access public places because the COVID-19 status of the users are checked before encountering other users and entering public places.
The KDC also does not issue new certificates for these users until they are removed from the lists of infected and suspected (close-contact) users. 
%Also, the user needs to authenticate himself/herself to the KDC for the renewing process of keys.
%To participate in the system, he/she needs to get a proof from a health authority that his/her status turns to -ve. 
%This proof is a signature from the health authority, which cannot be forged without having the health authority private key.

\vspace{2mm}\textit{\textbf{Preposition 5.} Our system is resilient against false reporting (or panic) attack that aims to falsely classify victim users as close contacts.
}% panic attack

\textbf{Proof.} 
In our system, each proof-of-contact (or proof-of-visiting) message includes a signature from the close-contact user (or public place), so to report false contacts to the health authority, the attacker should forge the signature of the victim user. 
This is impossible in our system without knowing the private key of the victim user. 
It is also impossible to compute the private key from the public key of the victim user under the known difficulty of computing the elliptic curve discrete logarithm \cite{ecdsad}. 
In addition, since the signatures of the proof-of-contact messages have a time stamp and the public key of the other user, attackers cannot use them for different users or different times, i.e., if a contact happened more than 14 days ago, attackers cannot alter the time and claim that the contact is recent.
Thus, adversaries cannot falsely claim that they have encountered users.  

% The health authority will verify the signature of $\sigma(time, location, Tmp_{\mathcal{PK}_U_{_j}})$ in any received message $M_U_{_i}=(Tmp_{\mathcal{PK}_U_{_i}},\sigma(time, location, Tmp_{\mathcal{PK}_U_{_j}}),time,\\location)$.
%The signature is used here to link $T_{\mathcal{PK}_{U_i}}$ and $T_{\mathcal{PK}_{U_j}}$ together to  prove that the two users have met in a specific time frame and a certain location (proof of contact). 
% Then it will check the message's integrity by comparing the time and location within the content of the signature $\sigma(time, location, Tmp_{\mathcal{PK}_U_{_j}})$ with the appended time and location in $M_U_{_i}$. If the time and location are the same, the integrity of the message is achieved. Otherwise, the message will be discarded as it may have been tampered with.
% % T_{\mathcal{PK}_{U_i}}

\vspace{2mm}\textit{\textbf{Preposition 6.} The process of updating COVID-19 status and zone’s status is transparent and verifiable. 
} % Transparency

\textbf{Proof.} 
Blockchain brings transparency to our system by not trusting a single entity (health authority) to update COVID-19 status and zone’s status. 
In fact, all transactions (proof-of-contact and proof-of-visiting messages), the updated lists of infected and close-contact users, and the updated zone categorization are verified by all the Blockchain nodes, and the majority of the validators should approve them. 
%One validator is elected as a leader based on the Raft consensus algorithm. 
As explained earlier, the leader health authority is responsible for creating blocks containing two Bloom filters for the COVID-19 status of the users and another blocks containing the updated status of zones. 
These blocks are broadcasted to the Blockchain network and the leader needs to get the votes of the majority of the validators before appending the blocks to the corresponding ledgers.
Therefore, a malicious health authority cannot manipulate the lists and the zone categorization. 
\end{enumerate}
\end{comment}

\section{Performance Evaluation}
\label{evaluate}

% {\color{red} 

% (1) The evaluation focuses on two phases, the \textit{contact} phase and the \textit{ tracing} phase.     Why?   We evaluate the system not two phases I removed this sentence 

% (2) It looks a weakness that we do not evaluate the overhead on the Blockchain in terms of on chain storage, overhead of validation of transactions (with and without aggregation), time needed to check the status of a user 

% so can we divide this section into two subsections for (1) Contact and Tracing phases (2) Blockchain   
% Are there other phases we do not evaluate?

% "In the \textit {contact} phase, the message sent from one user to another after mutual authentication and confirmation of -ve health"
% We do not do it this way -- you should evaluate the size fo each message exchanges in the protocols and refer to the figures you can also number of name messages so that we can refer to them easily here 

% }

In this section, the performance of our system is evaluated and compared to other systems.

\subsection{Experimental setup and performance metrics }
The signature scheme used in our system is based on the elliptic curve cryptography \cite{bos2014elliptic}. In our evaluations, we use 224-bit elliptic curve and SHA-224 hash function. 
Thus, the sizes of public key, private key, hash value, and signature are given in Table \ref{tab:size-offer}. 
Moreover, the sizes of other data items used in our system, including random number, location, and time are also given in in Table \ref{tab:size-offer}. In our experiments, we have used the Charm Python library \cite{akinyele2013charm} to calculate the time required to implement the different cryptographic  operations including bilinear  pairing (\textsl{Pairing}), hashing (\textsl{Hash}), addition (\textsl{Add}), multiplication (\textsl{Mul}), and exponentiation (\textsl{Exp}). The experiments have been conducted on Raspberry Pi 3 device with 1.2 GHz processor and 1 GB RAM. 
The time required to run the different cryptographic operations are given in Table \ref{cost1}. 

For the performance metrics, we evaluate our system in terms of communication, computation, and storage overheads.
\begin{itemize}
    \item \textbf{Communication overhead}. The sizes of the messages exchanged between the system entities. 
    \item \textbf{Computation overhead.} The time required to perform the steps of the system phases. %a  
    \item \textbf{Storage overhead.} The memory space required to store the data of our system. 
\end{itemize}
\begin{table}[!t]
\center
\caption{Sizes of the data used in our system. \label{tab:size-offer}}
\begin{tabular}{|c|c|}
\hline 
Data  & Size (bytes) \tabularnewline
%\hline 
%\hline 
%Real identity ($ID_j$) & 6  & Available services index ($S$) & 2\tabularnewline
%\hline 
%Certificate $Cert_U_{_i}$ & 72 & Random Number ($R$) & 4\tabularnewline
\hline \hline
Random number & 5\tabularnewline \hline
Private key  & 28\tabularnewline \hline
Public key  & 29 \tabularnewline \hline
Public key certificate & 93 \tabularnewline \hline
Hash & 28 \tabularnewline \hline
Signature & 56 \tabularnewline \hline
Location  & 6  \tabularnewline \hline
Time  & 8 \tabularnewline \hline 
% Zone ID  & 2 \tabularnewline \hline 
% Number of infected users & 2 \tabularnewline \hline 
% Number of suspected users & 2 \tabularnewline \hline 
\end{tabular}
\end{table}
%%%%%%%%%%%%%%%%%%%%%%%%%%%%%%%%%%%%%%%%%%%%%%%%%%%%%%%%%%%%%%%%%%%%%%%%%%%%%%%%%%%%%%%%%%%%%%
\begin{table}[!t]
		\centering
		\caption{Computation times of the cryptographic operations.}
		\begin{tabular}{|c|c|}
		\hline
		Cryptographic operation 	&   Time (ms)  \\\hline\hline
		\textsl{Pairing}  & 3.139    \\ \hline
		\textsl{Hash} & 0.058     \\ \hline	
		\textsl{Add} &  0.000227     \\ \hline	
		\textsl{Mul} & 0.000269      \\ \hline
		\textsl{Exp} &  0.334    \\ \hline
		\end{tabular}
		
		\label{cost1}
	\end{table}
%%%%%%%%%%%%%%%%%%%%%%%%%%%%%%%%%%%%%%%%%%%%%%%%%%%%%%%%%%%%%%%%%%%%%%%%%%%%%%%%%%%%%%%%%%%%%%

\subsection{Communication Overhead}
\subsubsection{\textbf{Contact Phase}}
% The communication overhead is measured in terms of the size of the transmitted messages in bytes. 
% The messages that we are going to measure their sizes are:
% \begin{itemize}
% \item The message sent from from one user to another in the \textit{ contact} phase.
% \item The message sent from an infected user to a health authority in the \textit{tracing} phase.
% %\item The message between health authority and Blockchain in the status update phase.
% %\item The message between the health authority and KDC in the status update phase.
% \end{itemize}
% In this subsection, we evaluate the communication, computation, and storage overheads of the \textit{contact} phase.

% \textbf{Communication overhead.} 
Fig. \ref{fig:Contact-phase} shows the exchanged messages between the system entities involved in the \textit{tracing} phase. From Table \ref{tab:size-offer}, we can calculate the sizes of these messages as given in Table \ref{com:contact}. 
The table demonstrates that our system requires exchanging small messages in the \textit{contact} phase. 
%In step  \raisebox{.5pt}{\textcircled{\raisebox{-.9pt}  {4}}}, each user checks the COVID-19 status of the other user by sending his temporary public key to the Blockchain. The Blockchain responds with a 2-bit response represents the status of the user (00: ``Infected'', 01: ``Suspected, 11: ``Not founed''), which is negligible compared to the size of the public key. 

\begin{table}[!t]
\caption{Sizes of the messages exchanged in the \textit{contact} phase.} \label{com:contact}
\centering
\renewcommand{\arraystretch}{1.3}
\resizebox{\columnwidth}{!}{%
\scriptsize
\begin{tabular}{|c|c|c|c|c|c|c|c|}
\hline
Message      & \raisebox{.5pt}{\textcircled{\raisebox{-.9pt}  {1}}}   & \raisebox{.5pt}{\textcircled{\raisebox{-.9pt}  {2}}} & \raisebox{.5pt}{\textcircled{\raisebox{-.9pt}  {3}}} & \circleded{\tiny4.1} & \circleded{\tiny4.2} & \raisebox{.5pt}{\textcircled{\raisebox{-.9pt}  {5}}} & \raisebox{.5pt}{\textcircled{\raisebox{-.9pt}  {6}}} \\ \hline
Size (bytes) & 5 & 154 & 149 & 29 & 29 & 99 & 99 \\ \hline
\end{tabular}%
}
\end{table}

% \textbf{Storage overhead}. Upon the completion of the \textit{contact} phase, the user stores the tuple  ($M_i, \sigma_i, Cert_i$) in his/her smart phone. The size of this tuple is $43+56+93=192$ bytes. Thus, after executing the \textit{contact} phase $\mathcal{M}$ times, the user should store ($\mathcal{M}\times192$) bytes. However, with the aggregate signature, the user only stores one signature ($\sigma_agg$) instead of $\mathcal{M}$ signatures. Thus, the storage overhead after executing the \textit{contact} phase $\mathcal{M}$ times is ($\mathcal{M}\times136+56$) bytes.

%%%%%%%%%%%%%%%%%%%%%%%%%%%%%%%%%%%%%%%%%%%%%%%%%%%%%%%%%%%%%%%%%%%%%%%%%%%%%%%%%%%%%%%%%%%%%%%%%%%%%%%
\subsubsection{\textbf{Tracing Phase}}
% In this subsection, we evaluate the communication and computation overheads of the \textit{tracing} phase.
% \begin{table}[!t]
% \caption{Sizes of the messages exchanged in the \textit{tracing} phase.} \label{com:tracing}
% \centering
% \renewcommand{\arraystretch}{1.3}
% \resizebox{\columnwidth}{!}{%
% \scriptsize
% \begin{tabular}{|c|c|c|c|c|c|}
% \hline
% Message      & \raisebox{.5pt}{\textcircled{\raisebox{-.9pt}  {1}}}   & \raisebox{.5pt}{\textcircled{\raisebox{-.9pt}  {2}}} & \raisebox{.5pt}{\textcircled{\raisebox{-.9pt}  {3}}} & \raisebox{.5pt}{\textcircled{\raisebox{-.9pt}  {4}}} & \raisebox{.5pt}{\textcircled{\raisebox{-.9pt}  {5}}} \\ \hline
% Size (bytes) & 5 & 154 & 149 & - & ($136\mathcal{N}+56\mathcal{M}$)  \\ \hline
% \end{tabular}%
% }
% \vspace{.5ex}

% Note:\hspace{.5ex} - means that the message size is negligible.
% \end{table}

% \textbf{Communication overhead.} 

% Fig. \ref{fig:Tracing_phase} shows the exchanged messages between the system entities involved in the \textit{contact} phase. 
% From Table \ref{tab:size-offer}, we can calculate the sizes of these messages as given in Table \ref{com:tracing}.
% In step \raisebox{.5pt}{\textcircled{\raisebox{-.9pt}  {5}}},
In this phase, the infected user uploads proof-of-contact and proof-of-visiting messages collected in the last 14 days ($\mathcal{N}$) as groups of $\mathcal{M}$ messages, where these groups are uploaded at different times. The size of each group is ($\mathcal{M}\times 136+56$) bytes. Thus, the communication overhead 
% of step \raisebox{.5pt}{\textcircled{\raisebox{-.9pt}  {5}}} 
is ($\frac{\mathcal{N}}{\mathcal{M}}\times(\mathcal{M}\times 136+56)$) = ($136\mathcal{N}+56\mathcal{M}$) bytes. 
Without using the signature aggregation technique, the overhead would be $192\mathcal{N}$.
%After receiving and verifying the uploaded messages from the infected user, 
The health authority broadcasts these messages to the other health authorities in the Blockchain network nodes (validators) to verify. %to be used in updating the Blockchain ledger in the \textit{status} update phase.
Thus, the communication overhead on the Blockchain validators is also ($136\mathcal{N}+56\mathcal{M}$) bytes. 

\begin{figure}
    \centering
    \includegraphics[width=\columnwidth]{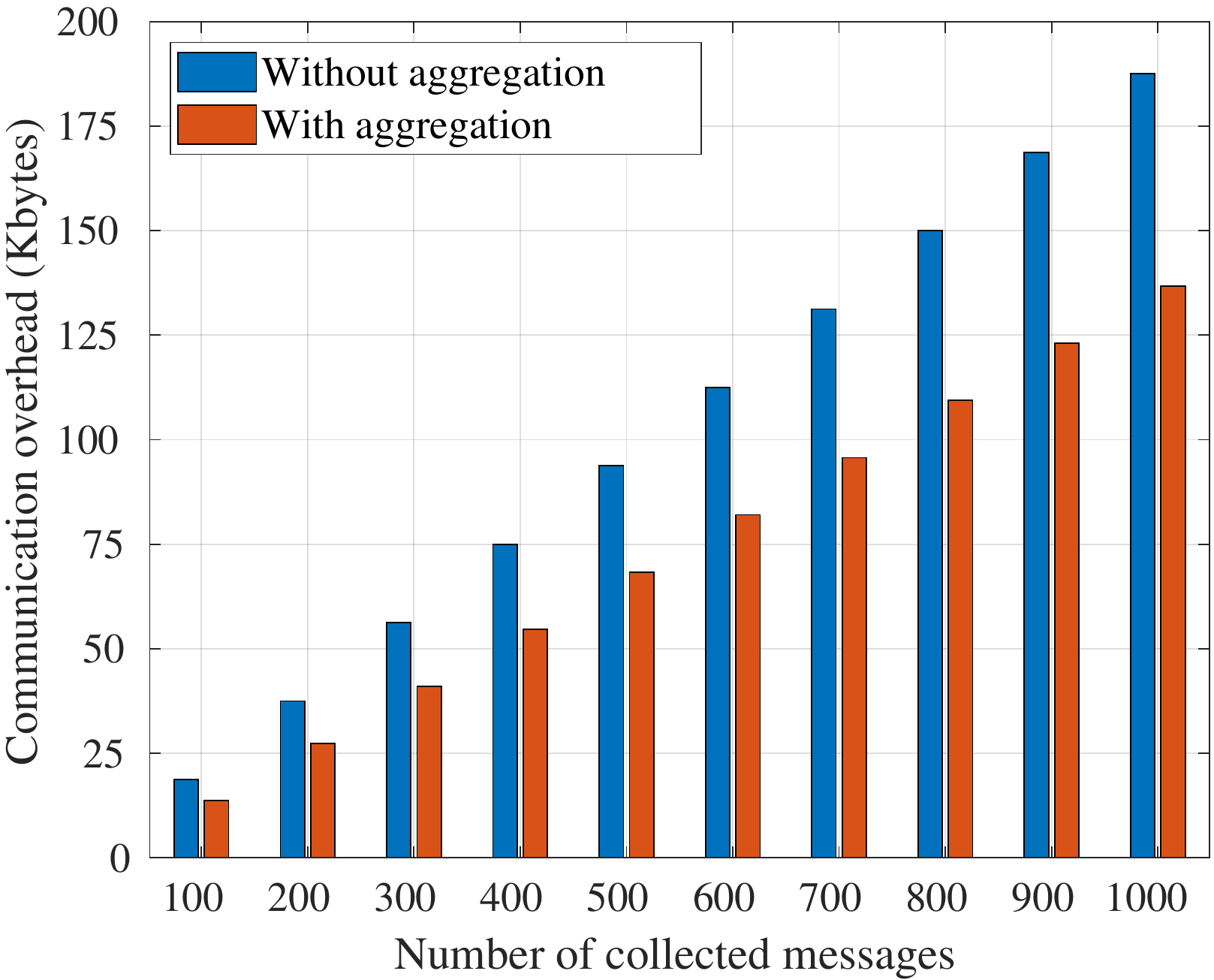}
    \caption{The communication overhead on the Blockchain validators with and without using aggregation.}
    \label{fig:wwcom}
\end{figure}

To illustrate the importance of using signature aggregation technique, Fig. \ref{fig:wwcom} compares between the communication overhead on the Blockchain validators with and without using aggregation at different numbers of messages in case $\mathcal{M}=\mathcal{N}/14$. 
The figure shows that the signature aggregation technique considerably reduces the communication overhead. 
For example, when the number of messages is $300$, the communication overhead with aggregation is only $41$Kbytes, while it is $56$Kbytes without aggregation. 
In other words, \textit{using the signature aggregation technique reduces the communication overhead by $27\%$}.

%%%%%%%%%%%%%%%%%%%%%%%%%%%%%%%%%%%%%%%%%%%%%%%%%%%%%%%%%%%%%%%%%%%%%%%%%%%%%%%%%%%%%%%%%%%%%%%%%%%%%%%%%
\subsubsection{\textbf{Status Update/Check Phase}} \label{filterSize}
\begin{table}[!t]
\caption{Sizes of the messages exchanged in the \textit{status update/check} phase.} \label{com:status}
\centering
\renewcommand{\arraystretch}{1.3}
\resizebox{\columnwidth}{!}{%
\large
\begin{tabular}{|c|c|c|c|c|c|c|}
\hline
Message      & \raisebox{.5pt}{\textcircled{\raisebox{-.9pt}  {1}}}   & \raisebox{.5pt}{\textcircled{\raisebox{-.9pt}  {2}}} & \raisebox{.5pt}{\textcircled{\raisebox{-.9pt}  {3}}} & \raisebox{.5pt}{\textcircled{\raisebox{-.9pt}  {4}}} & \circleded{\small5.1} & \circleded{\small5.2}\\ \hline
Size (bytes) & ($29\mathcal{V}$) & ($29\mathcal{V}$) & $(10N_1+10N_2)/8$ & (6$\mathcal{Z}$) & 29& 2  \\ \hline
\end{tabular}%
}
\vspace{.5ex}

Note:\hspace{.5ex} - means that the message size is negligible.
\end{table}
% The collected messages by each health authority along with the test results are broadcasted to the other health authorities in the Blockchain network, where one health authority is elected as a leader to update the COVID-19 status and the zones status ledger.  
%One of the Blockchain validators is elected as a leader to update the Blockchain ledgers and

\begin{figure}
    \centering
    \includegraphics[width=\columnwidth]{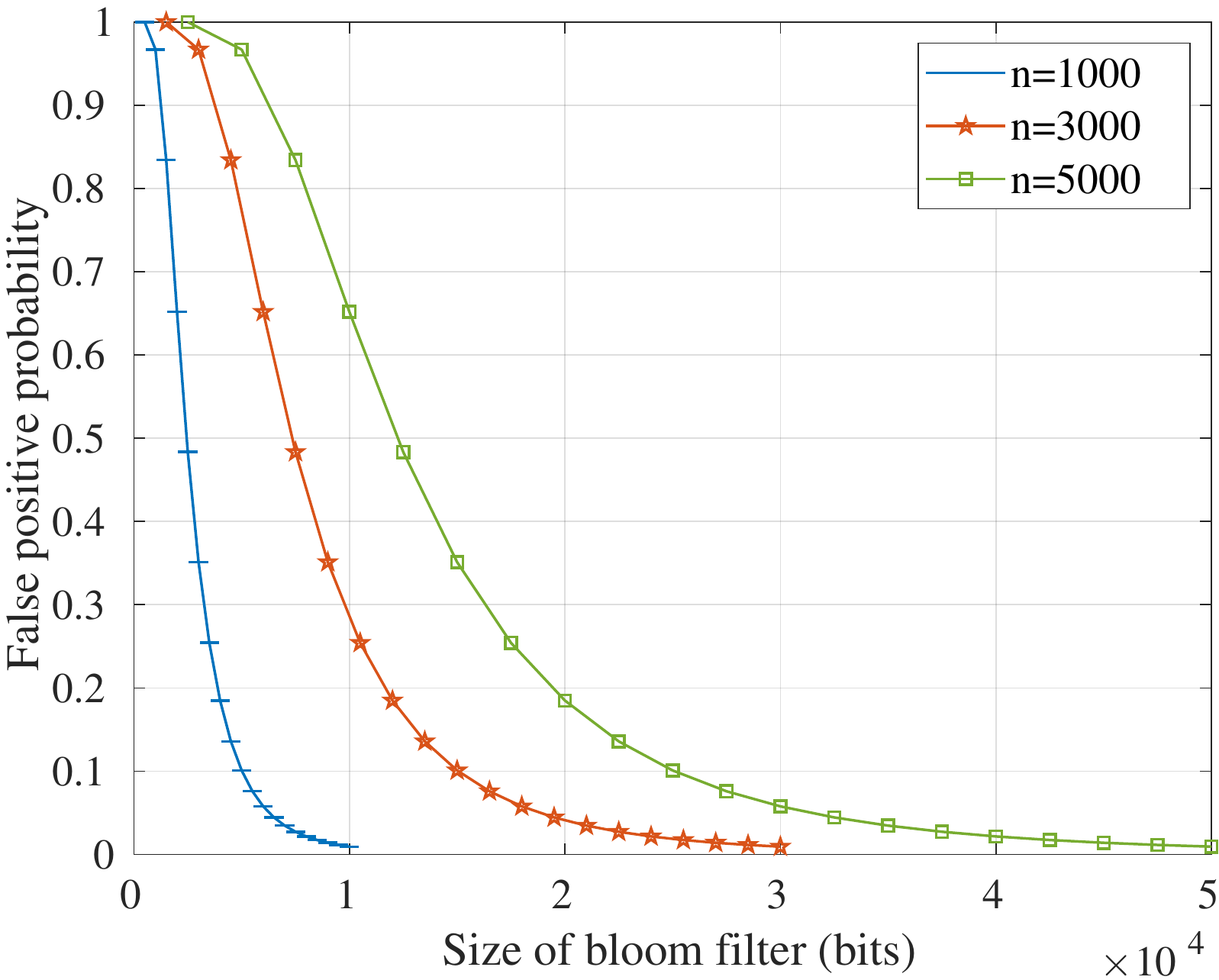}
    \caption{The relation between the false positive probability and the Bloom filter size at different numbers of public keys.}
    \label{falsepositive}
\end{figure}

\vspace{2mm}
Fig. \ref{fig:StatusUpdateP} shows the exchanged messages between the system entities in the \textit{status update/check} phase. From Table \ref{tab:size-offer}, we can calculate the sizes of these messages as given in Table \ref{com:status}. 
The communication overhead of step \raisebox{.5pt}{\textcircled{\raisebox{-.9pt}  {1}}} or \raisebox{.5pt}{\textcircled{\raisebox{-.9pt}  {2}}} is ($29\mathcal{V}$) bytes, where $\mathcal{V}$ is the number of keys with expired certificates. 
In step \raisebox{.5pt}{\textcircled{\raisebox{-.9pt}  {3}}}, the leader broadcasts two updated Bloom filters to the Blockchain followers. Thus, the communication overhead of step \raisebox{.5pt}{\textcircled{\raisebox{-.9pt}  {3}}} depends on the sizes of the Bloom filters. As explained in Section \ref{BloomFilter}, there is a probability of false positive when we check if an element is stored in the filter. According to Eq. \ref{falsepropab}, this probability depends on the number of hash functions ($k$), the size of the filter in bits ($m$), and the number of elements stored in the filter ($n$). 
In our system, we use five hash functions ($k=5$) and adaptive filter size that depends on the number of elements stored in the filter to keep the false positive probability low. 
Fig. \ref{falsepositive} illustrates the relation between the false positive probability and the size of the filter at different numbers of stored elements. 
We can observe from Fig. \ref{falsepositive} that the larger the size of the filter, the lower the false positive probability. We can also observe that the false positive probability is close to zero when the filter size is ten times the number of elements. Thus, the communication overhead of step \raisebox{.5pt}{\textcircled{\raisebox{-.9pt}  {3}}} is approximately equal to ($(10N_1+10N_2)/8$) bytes, where $N_1$ and $N_2$ are the number of infected and suspected users, respectively.

In step \raisebox{.5pt}{\textcircled{\raisebox{-.9pt}  {4}}}, the leader broadcasts the zones's IDs and their updated status to the Blockchain followers. 
Thus, the communication overhead of step \raisebox{.5pt}{\textcircled{\raisebox{-.9pt}  {4}}} is approximately equal to (6$\mathcal{Z}$) bytes, where $\mathcal{Z}$ is the number of the zones and each of zone's ID, number of infected users, and number of suspected users and the status is represented by $2$ bytes. 
%Note that the zone's status is just represented with $2$ bits ((00: ``Red'', 01: ``Orange, 11: ``Green'')), which is negligible. 
In step \raisebox{.5pt}{\textcircled{\raisebox{-.9pt}  {5}}}, a non-infected user can check his updated COVID-19 status by sending his temporary public key to the Blockchain or check a zone status by sending the zone ID to the Blockchain. In case that the non-infected user checks the COVID-19 status, the communication overhead of step \circleded{\scriptsize5.1} is approximately equal to $29$ bytes. In case that the non-infected user checks the zone status, the communication overhead of step \circleded{\scriptsize5.2} is approximately equal to $2$ bytes because the user sends a 2-byte zone ID to the Blockchain. 

%%%%%%%%%%%%%%%%%%%%%%%%%%%%%%%%%%%%%%%%%%%%%%%%%%%%%%%%%%%%%%%%%%%%%%%%%%%%%%%%%%%%%%%%%%%%%%%%%%%%%%%%
\begin{table}[!t]
\caption{Sizes of the messages exchanged in the \textit{public places access control} phase.} \label{com:places}
\centering
\renewcommand{\arraystretch}{1.1}
\resizebox{0.7\columnwidth}{!}{%
% \tiny
\begin{tabular}{|c|c|c|c|c|c|}
\hline
Message      & \raisebox{.5pt}{\textcircled{\raisebox{-.9pt}  {1}}}   & \raisebox{.5pt}{\textcircled{\raisebox{-.9pt}  {2}}} & \raisebox{.5pt}{\textcircled{\raisebox{-.9pt}  {3}}} & \raisebox{.5pt}{\textcircled{\raisebox{-.9pt}  {4}}} & \raisebox{.5pt}{\textcircled{\raisebox{-.9pt}  {5}}} \\ \hline
Size (bytes) & 5 & 154 & 149 & 29 & -  \\ \hline
\end{tabular}%
}
\vspace{.5ex}

Note:\hspace{.5ex} - means that the message size is negligible.
\end{table}

\vspace{2mm}
\subsubsection{\textbf{Public Places Access Control Phase}}
% \textbf{Communication overhead.} 
Fig. \ref{fig:COVID-19StatusVerification_phase} shows the exchanged messages between the system entities involved in the \textit{Public Places Access Control} phase. 
From Table \ref{tab:size-offer}, we can calculate the sizes of these messages and the results are given in Table \ref{com:places}. 
Overall, we can conclude that our system requires exchanging small-size messages in this phase. 

%%%%%%%%%%%%%%%%%%%%%%%%%%%%%%%%%%%%%%%%%%%%%%%%%%%%%%%%%%%%%%%%%%%%%%%%%%%%%%%%%%%%%%%%%%%%%%%%%%%%%%%%%%%%%%%%%%%%%%%%%%%%%%%%%%%%%%%%%%%%%%%%%%%%%%%%%%%%%%%%%%%%%%%%%%%%%%%%%%%%%%%%%%%%%%%%%%%%%%%%%%%%%%%%%

\subsection{Computation Overhead}
\subsubsection{\textbf{Contact Phase}}
\begin{table}[!t]
\caption{Computation times of the steps of the \textit{contact} phase.} \label{cop:contact}
\centering
\renewcommand{\arraystretch}{1.3}
\resizebox{\columnwidth}{!}{%
\scriptsize
\begin{tabular}{|c|c|c|c|c|c|c|c|}
\hline
Step      & \raisebox{.5pt}{\textcircled{\raisebox{-.9pt}  {1}}}   & \raisebox{.5pt}{\textcircled{\raisebox{-.9pt}  {2}}} & \raisebox{.5pt}{\textcircled{\raisebox{-.9pt}  {3}}} & \circleded{\tiny4.1} & \circleded{\tiny4.2} & \raisebox{.5pt}{\textcircled{\raisebox{-.9pt}  {5}}} & \raisebox{.5pt}{\textcircled{\raisebox{-.9pt}  {6}}} \\ \hline
Time (msec) & - & 12.6 & 12.6 & 0.29 & 0.29 & 6.3 & 6.3 \\ \hline
\end{tabular}%
}
\vspace{.5ex}

Note:\hspace{.5ex} - means that the computation time is negligible.
\end{table}
% \textbf{Computation overhead}. 
Using the computational times in Table \ref{cost1}, we can compute the computation times of the steps of the \textit{contact} phase and the results are given in Table \ref{cop:contact}. 
In step \raisebox{.5pt}{\textcircled{\raisebox{-.9pt}  {2}}} or \raisebox{.5pt}{\textcircled{\raisebox{-.9pt}  {3}}}, 
one user computes a signature and the other user verifies the certificate (by verifying the KDC's signature) and the signature of the user. 
According to Eq. \ref{hash}, the computation of a signature requires one \textsl{Hash} and one \textsl{Mul} operations; thus, the computation time is $0.058+0.000269 \approx 0.0583$ msec. 
According to Eq. \ref{VER}, the verification of a signature requires two \textsl{Pairing} operations; thus, the verification time is $2\times3.139=6.278$ msec. 
The computation overhead of step \raisebox{.5pt}{\textcircled{\raisebox{-.9pt}  {2}}} or \raisebox{.5pt}{\textcircled{\raisebox{-.9pt}  {3}}} is $0.0583+2\times6.278\approx12.6$ msec. In step \raisebox{.5pt}{\textcircled{\raisebox{-.9pt}  {4}}}, for the Blockchain to return the COVID-19 status of the user, it is required to compute $k$ hash functions to search for the user's key in the Bloom filters. In our system, we use $k=5$; thus, the computation overhead of step \raisebox{.5pt}{\textcircled{\raisebox{-.9pt}  {4}}} is $5\times0.058=0.29$ msec. In step \raisebox{.5pt}{\textcircled{\raisebox{-.9pt}  {5}}} or \raisebox{.5pt}{\textcircled{\raisebox{-.9pt}  {6}}}, one user sends a proof-of-contact message signed with his/her temporary private key and the other user verifies the signature; thus, the computation overhead of \raisebox{.5pt}{\textcircled{\raisebox{-.9pt}  {5}}} or \raisebox{.5pt}{\textcircled{\raisebox{-.9pt}  {6}}} is $0.0583+6.278\approx6.3$ msec.

%%%%%%%%%%%%%%%%%%%%%%%%%%%%%%%%%%%%%%%%%%%%%%%%%%%%%%%%%%%%%%%%%%%%%%%%%%%%%%%%%%%%%%%%%%%%%%%%%%%%%%%%
\subsubsection{\textbf{Tracing Phase}}
% % \textbf{Computation overhead}.
% \begin{table}[!t]
% \caption{Computation times of the steps of the \textit{tracing} phase.} \label{cop:tracing}
% \centering
% \renewcommand{\arraystretch}{1.4}
% \resizebox{\columnwidth}{!}{%
% \small
% \begin{tabular}{|c|c|c|c|c|c|}
% \hline
% Step      & \raisebox{.5pt}{\textcircled{\raisebox{-.9pt}  {1}}}   & \raisebox{.5pt}{\textcircled{\raisebox{-.9pt}  {2}}} & \raisebox{.5pt}{\textcircled{\raisebox{-.9pt}  {3}}} & \raisebox{.5pt}{\textcircled{\raisebox{-.9pt}  {4}}} & \raisebox{.5pt}{\textcircled{\raisebox{-.9pt}  {5}}} \\ \hline
% Time (msec) & - & 12.6 & 12.6 & 0.0583 & $\frac{\mathcal{N}}{\mathcal{M}}\times3.139(\mathcal{M}+1)$ \\ \hline
% \end{tabular}%
% }
% \vspace{.5ex}

% Note:\hspace{.5ex} - means that the computation time is negligible.
% \end{table}
%From Table \ref{cost1}, we can compute the computation times of the steps of the \textit{tracing} phase.
% as given in Table \ref{cop:tracing}. 
% In step \raisebox{.5pt}{\textcircled{\raisebox{-.9pt}  {4}}}, the health authority signs the user's temporary public key; thus the computation overhead of step \raisebox{.5pt}{\textcircled{\raisebox{-.9pt}  {4}}} is $0.0583$ msec.
% In step \raisebox{.5pt}{\textcircled{\raisebox{-.9pt}  {5}}},
In this phase, after receiving $\mathcal{M}$ messages from the infected user, the health authority verifies only one aggregate signature ($\sigma_{agg}$) instead of verifying $\mathcal{M}$ individual signatures. 
According to Eq. \ref{aggregate}, the verification of $\sigma_{agg}$ requires $\mathcal{M}+1$ \textsl{Pairing} operations; thus, the verification time is $3.139(\mathcal{M}+1)$ msec. The total computation overhead
% of step \raisebox{.5pt}{\textcircled{\raisebox{-.9pt}  {5}}}
is ($\frac{\mathcal{N}}{\mathcal{M}}\times3.139(\mathcal{M}+1)$) msec. On the other hand, the computation overhead 
% of step \raisebox{.5pt}{\textcircled{\raisebox{-.9pt}  {5}}} 
would be $\mathcal{N}\times2\times3.139$ msec without using the signature aggregation technique, where the verification of each signature requires two \textsl{Pairing} operations according to Eq. \ref{VER}.

% \textbf{Storage overhead}. After uploading the $\mathcal{N}$ messages collected by the user in the last 14 days, the storage overhead of the \textit{tracing phase} on the health authority is  

\begin{figure}
    \centering
    \includegraphics[width=\columnwidth]{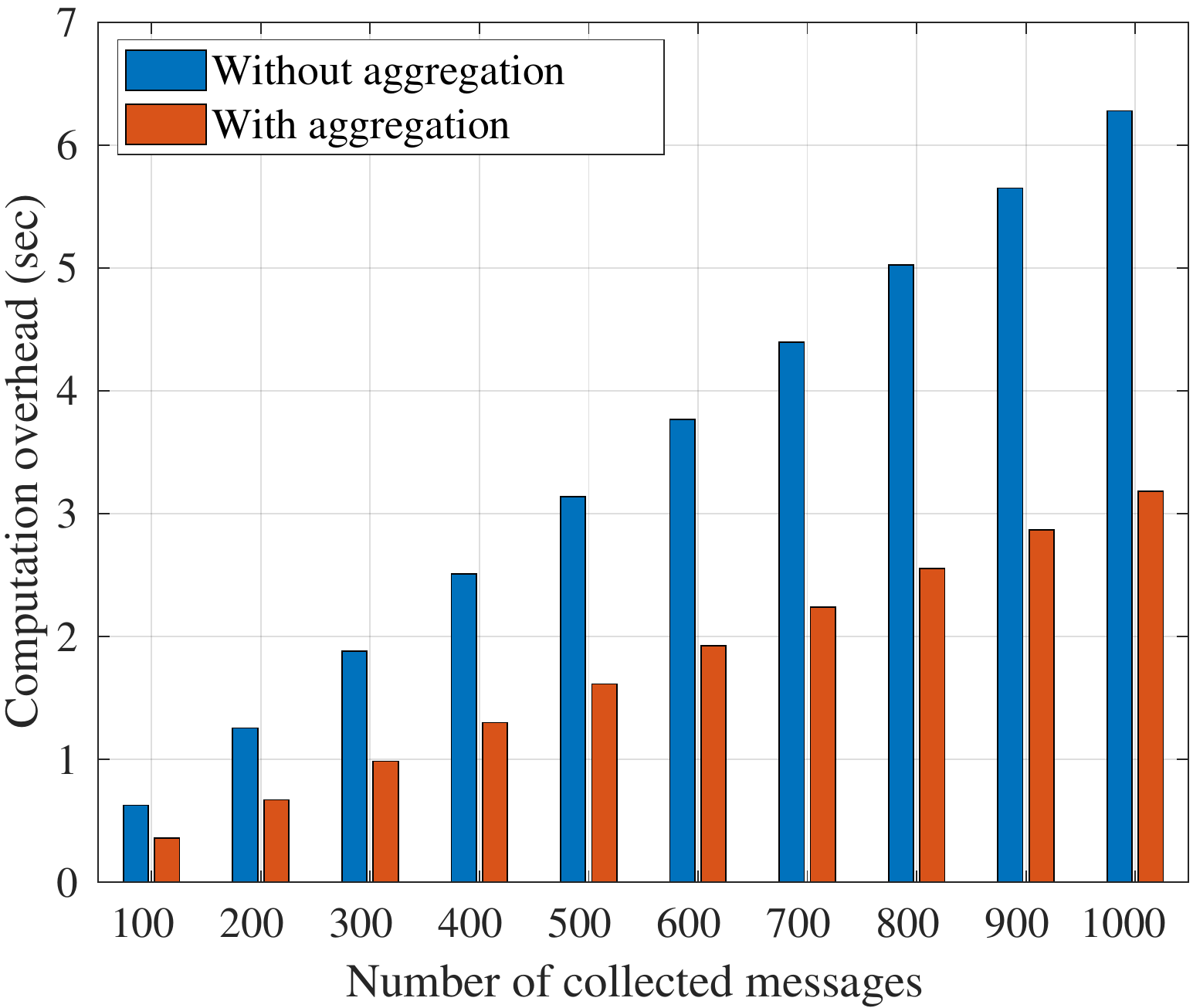}
    \caption{The computation overhead on the Blockchain validators with and without using aggregation.}
    \label{fig:wwcop}
\end{figure}

After verifying the messages uploaded from an infected user, the health authority broadcasts these messages to the other Blockchain validators to verify. %to be used in updating the Blockchain ledger in the \textit{status} update phase.
Thus, the computation overhead on each Blockchain validator is also ($\frac{\mathcal{N}}{\mathcal{M}}\times3.139(\mathcal{M}+1)$) msec. To evaluate the reduction in the computation overhead due to using the signature aggregation technique, in Fig. \ref{fig:wwcop}, we compare between the computation overhead on the Blockchain validators with and without using aggregation at different numbers of messages (in case $\mathcal{M}=\mathcal{N}/14$). 
The figure shows that the signature aggregation technique considerably reduces the computation overhead. For example, when the number of messages is $300$, the computation overhead with aggregation is approximately $1$ sec, which is nearly half the computation overhead without aggregation. In other words, \textit{using the signature aggregation technique results in about $48\%$ reduction in the computation overhead}.
%%%%%%%%%%%%%%%%%%%%%%%%%%%%%%%%%%%%%%%%%%%%%%%%%%%%%%%%%%%%%%%%%%%%%%%%%%%%%%%%%%%%%%%%%%%%%%%%%%%%%%%%
\subsubsection{\textbf{Status Update/Check Phase}}
\begin{table}[!t]
\caption{Computation times of the steps of the \textit{status update/check} phase.} \label{cop:status}
\centering
\renewcommand{\arraystretch}{1.4}
\resizebox{\columnwidth}{!}{%
\scriptsize
\begin{tabular}{|c|c|c|c|c|c|c|}
\hline
Step      & \raisebox{.5pt}{\textcircled{\raisebox{-.9pt}  {1}}}   & \raisebox{.5pt}{\textcircled{\raisebox{-.9pt}  {2}}} & \raisebox{.5pt}{\textcircled{\raisebox{-.9pt}  {3}}} & \raisebox{.5pt}{\textcircled{\raisebox{-.9pt}  {4}}} & \circleded{\tiny5.1} & \circleded{\tiny5.2}\\ \hline
Time (msec) & - & - & $0.29(N_1+N_2)$ & - & 0.29& - \\ \hline
\end{tabular}%
}
\vspace{.5ex}

Note:\hspace{.5ex} - means that the computation time is negligible.
\end{table}
In step \raisebox{.5pt}{\textcircled{\raisebox{-.9pt}  {3}}} of the \textit{status update/check} phase, for the leader to create a Bloom filter, he needs to calculate five \textsl{Hash} operations for each public key to be stored in the filter; thus, the computation overhead of step \raisebox{.5pt}{\textcircled{\raisebox{-.9pt}  {3}}} is $(5\times0.058\times(N_1+N_2)) = 0.29(N_1+N_2)$ msec.

%%%%%%%%%%%%%%%%%%%%%%%%%%%%%%%%%%%%%%%%%%%%%%%%%%%%%%%%%%%%%%%%%%%%%%%%%%%%%%%%%%%%%%%%%%%%%%%%%%%%%%%%%
\subsubsection{\textbf{Public Places Access Control Phase}}
\begin{table}[!t]
\caption{Computation times of the steps of the \textit{public places access control} phase.} \label{cop:places}
\centering
\renewcommand{\arraystretch}{1.1}
\resizebox{0.7\columnwidth}{!}{%
% \tiny
\begin{tabular}{|c|c|c|c|c|c|}
\hline
Step      & \raisebox{.5pt}{\textcircled{\raisebox{-.9pt}  {1}}}   & \raisebox{.5pt}{\textcircled{\raisebox{-.9pt}  {2}}} & \raisebox{.5pt}{\textcircled{\raisebox{-.9pt}  {3}}} & \raisebox{.5pt}{\textcircled{\raisebox{-.9pt}  {4}}} & \raisebox{.5pt}{\textcircled{\raisebox{-.9pt}  {5}}} \\ \hline
Time (msec) & - & 12.6 & 12.6 & 0.29 & - \\ \hline
\end{tabular}%
}
\vspace{.5ex}

Note:\hspace{.5ex} - means that the computation time is negligible.
\end{table}
Using Table \ref{cost1}, we can compute the computation times of the steps of the \textit{public places access control} phase, and the results are given in Table \ref{cop:places}. 
The table shows that the computation times of this phase are acceptable. 

%%%%%%%%%%%%%%%%%%%%%%%%%%%%%%%%%%%%%%%%%%%%%%%%%%%%%%%%%%%%%%%%%%%%%%%%%%%%%%%%%%%%%%%%%%%%%%%%%%%%%%%%%
%%%%%%%%%%%%%%%%%%%%%%%%%%%%%%%%%%%%%%%%%%%%%%%%%%%%%%%%%%%%%%%%%%%%%%%%%%%%%%%%%%%%%%%%%%%%%%%%%%%%%%%
\subsection{Storage Overhead}
\subsubsection{\textbf{Off-Chain Storage}}
Upon the completion of the \textit{contact} phase, the user stores the tuple ($M_i, \sigma_i, Cert_i$) in his/her smart phone as a proof-of-contact. The size of this tuple is $43+56+93=192$ bytes. 
Thus, for $\mathcal{M}$ proof-of-contact messages, the user stores ($192\mathcal{M}$) bytes. 
However, with the signature aggregation technique, the user only stores one signature ($\sigma_{agg}$) instead of $\mathcal{M}$ individual signatures. 
Thus, the storage overhead for $\mathcal{M}$ proof-of-contact (or proof-of-visit) messages is ($136\mathcal{M}+56$) bytes.
%%%%%%%%%%%%%%%%%%%%%%%%%%%%%%%%%%%%%%%%%%%%%%%%%%%%%%%%%%%%%%%%%%%%%%%%%%%%%%%%%%%%%%%%%%%%%%%%%%%%%%%%%

\subsubsection{\textbf{On-Chain Storage}}
The Blockchain network in our system stores two ledgers for the COVID-19 status and the zones status. 
Each ledger consists of blocks, where each block has a header and a body. Based on the Raft consensus algorithm, the header of each block contains the hash of its previous block in the ledger, timestamp, and the leader's signature. 
Thus, the block header is $100$ bytes. 
In the COVID-19 status ledger, each block body has two Bloom filters, and in Section \ref{filterSize}, we found that to make the false positive probability very low, the filter size should equal to $10n$ bits, where $n$ is the number of stored elements in the filter. 
Thus, the total size of the block in the COVID-19 status ledger is ($100+(10N_1+10N_2)/8$) bytes, where $N_1$ and $N_2$ are the number of infected and suspected users, respectively. In the zones status ledger, each block body has zones' IDs, numbers of infected and suspected users in these zones, and zones' status. Thus, the total size of the block in the zones status ledger is approximately ($100$+ 6$\mathcal{Z}$) bytes, where $\mathcal{Z}$ is the number of the zones.

\begin{figure}
    \centering
    \includegraphics[width=\columnwidth]{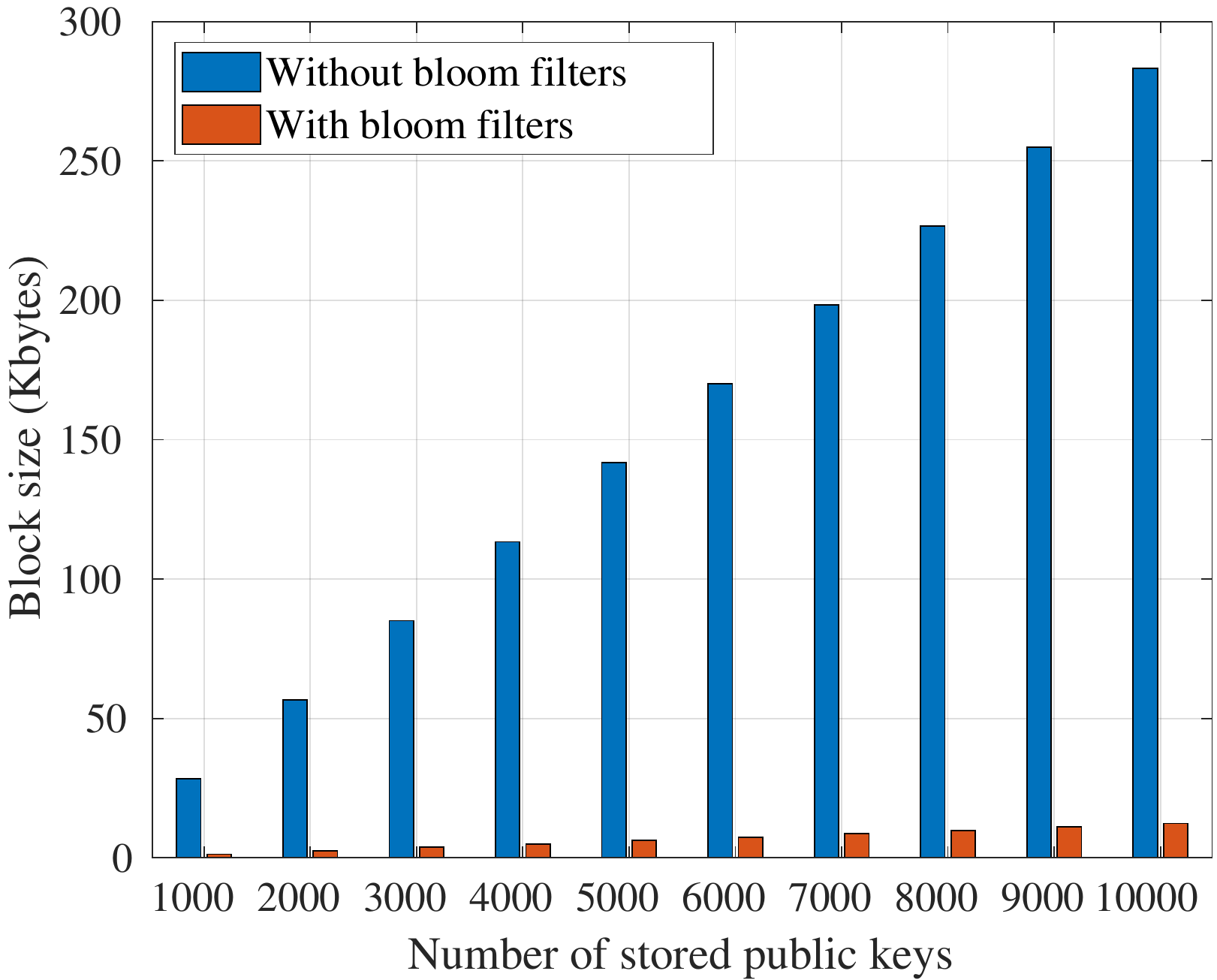}
    \caption{The COVID-19 status ledger's block size with and without using Bloom filters.}
    \label{storageOverhead}
\end{figure}

To evaluate the reduction in the storage space due to using Bloom filters, in Fig. \ref{storageOverhead}, we compare the block size of the COVID-19 status ledger with and without Bloom filters at different numbers of stored public keys ($N=N_1+N_2$). 
The figure shows that the block size is significantly reduced by using the Bloom filter. 
For example, if the number of public keys stored in the block is $7000$, the block size with Bloom filters is about $10$ Kbytes, while it is about $200$ Kbytes without Bloom filters. 
This indicates that \textit{the use of Bloom filters results in about $95\%$ reduction in the block size}. 
This is because without using Bloom filters, each public key needs $29$ bytes storage space instead of just $10$ bits if the Bloom filter is used.

\begin{table*}[!t]
\centering
\caption{Comparison between our system and the existing  infection control systems.}
\label{table:compare with other works}
\renewcommand{\arraystretch}{1.5}
\resizebox{\textwidth}{!}{%
%\hspace*{-1cm}
\begin{tabular}{|c|c|c|c|c|c|c|c|}

\hline
% \multicolumn{2}{|c|}\diagbox[width=12em, height=99]{\raisebox{4ex}{\hspace{20 pt}  Category}}{\raisebox{-11ex}{Papers} } \raggedright
% \diagbox[width=21em,height=29]{{
% \hspace{5 pt} Features}}{{\hspace{-15 pt} Systems} }
 \multicolumn{2}{|c|}{Comparison}&  ContactChaser \cite{wancontactchaser}  & EPIC \cite{altuwaiyan2018epic}&PHyCT \cite{jhanwarphyct}&CAUDHT \cite{brack2020decentralized}& \multicolumn{1}{p{1.8cm}|} { \centering \cite{torky2020covid}}  & Our system \\
%PHyCT \cite{}  & $\surd$ &  $\times$&  $\surd$ & $\surd$ &  $\times$ & $\surd$ & $\times$& $\times$ & $\times$ &$\times$  & $\surd$ & $\bigoplus$\\ \hline
\hline %\rule{0pt}{25pt}
\multirow{3}{*}{  Privacy of infected users}&  Protection against Identification attack  & $\surd$ &$\surd$ &$\surd$ &$\surd$ &$\times$& $\surd$\\ \cline{2-8}
 & Prevention of Social graph disclosure &$\surd$ & $\surd$ &$\times$ &$\surd$  &$\times$ & $\surd$ \\ \cline{2-8}
  & Protection against Tracking attack &  $\surd$&$\times$ &$\bigoplus$&$\bigoplus$ &$\times$ & $\surd$ \\  \hline
\multirow{2}{*}{ Privacy of close contacts}& Protection against Identification attack  &$\surd$ &$\surd$ &$\surd$ &$\surd$ &$\times$& $\surd$\\ \cline{2-8}
& Protection against Tracking attack &  $\bigcirc$ & $\times$ &$\bigoplus$&$\bigoplus$ &$\times$ & $\surd$ \\  \hline
\multirow{2}{*}{ Privacy of non-infected users}& Protection against Identification attack  &$\surd$ &$\surd$ &$\surd$ &$\surd$ &$\times$& $\surd$\\ \cline{2-8}
  &  Protection against Tracking attack &  $\surd$ & $\surd$&$\bigoplus$&$\bigoplus$ &$\times$& $\surd$ \\ \hline
\multicolumn{2}{|c|}{Prevention of False Reports}  &$\times$ &$\times$ &$\surd$ &$\surd$ &$\surd$& $\surd$\\  \hline
\multicolumn{2}{|c|}{Tracing \& Notification}   &  $\surd$ &$\surd$ &$\surd$ &$\surd$ &$\surd$ &$\surd$ \\ \hline  
\multicolumn{2}{|c|}{Digital Pass}   &$\times$ &$\times$ &$\times$ & $\times$&  $\times$& $\surd$ \\   \hline
\multicolumn{2}{|c|}{Recommendation of safe Places}   &$\times$ &$\bigoplus$ &$\times$ & $\times$&$\bigoplus$ &$\surd$ \\  \hline
% \multicolumn{2}{|c|}{Low Communication \& Computation}   &$\surd$ &$\times$&$\times$ &$\surd$ &$\times$ &$\surd$ &    \hline
\multicolumn{2}{|c|}{Low storage usage}   & $\times$ &$\times$ &$\times$ & $\times$& $\times$& $\surd$\\  \hline
\multicolumn{2}{|c|}{Blockchain-based}  & $\times$ &$\times$ &$\times$ & $\times$& $\surd$& $\surd$\\  \hline
\end{tabular}%
}
\vspace{3mm}Note:\hspace{.5ex} $\surd$: a realized feature,  $\times$: an unrealized feature, $\bigcirc$: partially realized feature, and $\bigoplus$: not considered.
%\vspace{-0.3cm}
\end{table*}

\section{Related Work}\label{chap:Literature Review} 

Since the outbreak of COVID-19 pandemic, several systems have been proposed in the literature to control the infection, especially through contact tracing. These systems can be categorized into centralized and decentralized based on the underlying architecture. In this section, we briefly explain these systems and discuss their limitations.
% , and then we compare our system with the state-of-the-art
% works.

\subsection{Centralized Systems  \label{sec:centralized-Solutions}}
Most of the existing contact tracing systems are based on a centralized architecture \cite{sun2020vetting}, where a single central authority is responsible for identifying and notifying the close contacts of infected individuals. 

% Singapore government has implemented an application (app) for contact tracing called traceTogather \cite{Tracetogether}. This app asks the users to enter their mobile numbers, and then it pushes pseudonym IDs signed with the Ministry of Health (MoH) private key. When users encounter each other, they exchange messages containing their current pseudonym IDs. Each user stores the received IDs locally accompanied with the receiving times and locations. If a user is diagnosed to have COVID-19, all the collected IDs of the infected user are shared with MoH so that it can identify and send messages to the probably infected users. However, the MoH can learn detailed information on the users' visited locations and social activities, which raises privacy concerns and may discourage users to use the system. 
% Another COVID-19 contact tracing app, called Aarogya Setu \cite{Aarogya}, has been released in India. The app identifies and notifies the close contacts via searching in a database of known infected users.
% However, the security and privacy analysis conducted in \cite{sun2020vetting} reveals that Arogya Setu is vulnerable to multiple attacks, including false reporting and identification.

%%%%%%%%%%%%%%%%%%%%%%%%%%%%%%%%%%%%%%%%%%%%%%%%%%%%%%%%%%%%%%%%%%%%%%%%%%%%%%%%%%%%%%%%%%%%%%%%%%%%%%%%% 
Wan et al. \cite{wancontactchaser} have proposed a contact tracing system, called ContactChaser, based on group signature cryptosystem. 
In ContactChaser, the health authority is the group manager that generates private keys for registered users. 
When two users encounter each others, they generate group signatures on the current time and location, and exchange these signatures with each other. If a user is tested positive for COVID-19, he/she uploads to the health authority all the signatures collected in the past 14 days. 
The authority identifies the close contacts by tracing the signers of the signatures. 
% To prevent the health authority from creating a social graph, ContactChaser introduces a new entity, called mixer, that shuffles the signatures received from multiple infected users before forwarding them to the health authority so that the authority cannot know the users who contacted each other.
However, panic attacks can be launched because a user does not sign on the public key of the other user when they meet to prevent the authority from creating a social graph, and thus attackers can collect signatures issued for different users and claim that they are their close contacts.  
Another drawback is that the health authority can learn the locations visited by the users and their activities because it knows the real identity of the signer and the time and location of the contact are signed.
%In group signature, users do not have public keys.

% ContactChaser claims to counter false reports by rejecting huge amounts of messages sent by an individual even if they were valid. However, this does not solve the problem because a malicious user can still send a small number of messages as a false report attempt and get accepted by the mixer. Moreover, some close contacts will not get notifications because their corresponding infected users have encountered so many users in the past 14 days.
%%%%%%%%%%%%%%%%%%%%%%%%%%%%%%%%%%%%%%%%%%%%%%%%%%%%%%%%%%%%%%%%%%%%%%%%%%%%%%%%%%%%%%%%%%%%%%%%%%%%%%%%%%
 
Altuwaiyan et al. \cite{altuwaiyan2018epic} have proposed a contact tracing sytem, called EPIC, based on homomorphic encryption. The idea is that when a user visits a place, he/she collects messages from wireless devices in this place. 
If a user is infected, he/she sends the messages to a central server that enables non-infected users to learn whether they were close enough to an infected user. 
The system uses homomorphic encryption scheme to learn whether the users were close enough from each other. 
However, EPIC is vulnerable to tracking attack because the server can know the locations visited by the infected users due to including the hashed identifiers of the wireless devices located in each place. 
%These identifiers could be used by the server to retrieve the places visited by the infected users.  
%  The privacy issue in this system is that the infected individual discloses location data to the server.
%  Location data could be used to classify places based on the number of infected users been in a place and reflect current infection levels; however, the authors did not include this function in their system.
Also, EPIC is vulnerable to panic attack, where a malicious user can get identifiers of places that he/she did not visit and upload them to the server.
% network scans that do not belong to him/her to spread panic by letting other users think that they may be at risk of having COVID-19.
Moreover, EPIC suffers from high computation, communication, and storage overheads due to using the inefficient homomorphic encryption scheme.

Jhanwar et al. \cite{jhanwarphyct} have proposed a contact tracing system, called PHyCT. In this system, each user generates a short-term secret seed. Then, the user computes two shares (broadcast share and authority share) from the seed using a 2-out-of-2 secret sharing algorithm. All seeds are kept locally in the user device. The user consistently uploads his/her identity encrypted with the secret seed, authority share, and hash of the broadcast share to a central authority database.
% Afterward, the user computes temporary identifiers using a Pseudo-random function on the secret seed and time. Each time a user encounters another user, he/she shares a message contains a temporary identifier and broadcast share. 
When users encounter each other, they exchange broadcast shares.
Once a user gets infected, he/she uploads all the secret seeds and collected messages to the central authority database to be used by other users to identify if they contacted the infected user. 
% Non-infected users periodically conduct a local search by recomputing temporary identifiers based on the uploaded seeds and check for a match.
Close-contact users must report to the health department to be tested. 
For non-compliant users who have not reported to the health, the central authority reveals their identities by hashing the broadcast share obtained from the collected messages and searching for the corresponding entry.
% Then, it reconstructs the secret seed by using the broadcast share and central authority share. Now, the central authority has the private key.
This makes PHyCT vulnerable to social graph disclosure because the authority can reveal the identities of both non-compliant and compliant users. 
\subsection{Decentralized Systems   \label{sec:decentralized-solutions}}
Decentralized systems depend on a decentralized architecture using distributed hash table (DHT) or Blockchain, where multiple entities are responsible for identifying and notifying the close contacts of infected individuals. 
% conducting the contact tracing process and informing close contacts of their possible infection in a distributed manner.

Brack et al. \cite{brack2020decentralized} have proposed a contact tracing system, called CAUDHT. 
The idea is that when users encounter each other, they exchange temporary IDs. 
If a user is tested positive for COVID-19, he/she blinds his/her IDs used during the last 14 days and sends them to the health authority to sign as a proof of infection. 
Then, the infected user unblinds the signatures received from the health authority to obtain valid signatures on his/her IDs.
To notify a close-contact user, the infected user computes a ciphertext by encrypting the signature on his/her ID used during the contact with the ID of the close-contact user. 
Then, the infected user creates a record on a distributed hash table (DHT) that consists of the close contact's ID and the ciphertext to notify the close contact user. 
However, this system suffers from a large storage overhead due to storing ciphertexts in the DHT.

Torky et al. \cite{torky2020covid} have proposed a Blockchain-based contact tracing system. 
The system consists of a Blockchain platform, mass-surveillance cameras, peer-to-peer (P2P) mobile application, and an infection verifier.
% In this system, infected users are represented by  regular expressions and their close contacts are represented by instances generated from the regular expressions. 
Once a user is tested positive for COVID-19, a regular expression is created for him/her and recorded in the Blockchain platform. 
Then, the Blockchain submits a tracking request to mass-surveillance cameras monitoring the places visited by the user to identify his/her close contacts. 
The close contacts receive messages containing their infection instances via the P2P mobile application.
% A close contact can send his infection instance to the infection verifier subsystem, where the regular expressions of the infected users are represented by as finite automata. The infection verifier subsystem verifies the the infection instance if it belongs to any of finite automata.
A close contact user forwards the received infection instances to the infection verifier subsystem to verify it.
However, this system is vulnerable to social graph disclosure and tracking due to using a mass-surveillance system. 
Also, mass-surveillance systems are not available in many countries and it is costly to deploy a system for controlling COVID-19 infections.

\subsection{Limitations and Contributions}
%In this subsection, we discuss the limitations of the existing infection control systems and the difference between our system and these systems.

The systems in \cite{wancontactchaser, altuwaiyan2018epic, jhanwarphyct} are centralized systems; thus, they suffer from the following issues. 
First, they are vulnerable to a single point of failure problem and DoS attack. 
Second, they lack transparency because the users have to trust a single central authority for performing the contact tracing without being able to validate the operations done by the server. 
On the other hand, very few decentralized contact tracing systems have been proposed \cite{brack2020decentralized, torky2020covid}, but they suffer from several limitations. 
In Table \ref{table:compare with other works},  we compare our system to the existing works in terms of privacy and security features, infection control services, and performance metrics.

% This makes the system implicitly breach non-infected users' privacy and opens the possibility for authorities to track all users.  
% This makes the system implicitly breach non-infected users' privacy and opens the possibility for authorities to track all users.   The authority can locate the highest infection probability cases which invade the privacy of those users. Moreover, the system included human-to-human virus transmission and human-to-place virus transmission but failed to preserve users' privacy while tracing such transmissions. It also sends warning messages to the user when he/she is close to someone or in a crowdy cluster of people. The warning message contains the identity of all individuals and their infection probability rates. This makes the system suffer from an identification attack as users know about others' COVID-19 status along with authorities. The idea of revealing the contagion sources' identity makes them live in fear of public stigmatization and social repercussions. 

Our main contributions can be summarized as follows. 
First, compared to the existing systems, our system is the only one that preserves the privacy of the infected users and their close contacts against identification, social graph disclosure, and tracking attacks, while thwarting false reporting attacks.
As explained earlier, although \cite{torky2020covid} utilizes the Blockchain, it is vulnerable to social graph disclosure and tracking due to using a mass-surveillance system. Also, mass-surveillance systems are not available in many countries and it is costly to deploy a system for controlling COVID-19 infections.
Second, most of the existing systems focus only on contact tracing, which is not enough to control the infection. Our infection control system not only considers contact tracing, but also records the COVID-19 status of the system users to be used as a digital pass to prevent infected and suspected users from accessing public places to prevent spreading the virus by them. 
Moreover, our system categorizes zones according to their infection level so that users can avoid visiting highly contaminated zones to avoid contracting the virus.
Finally, our system requires low computation and communication overheads and storage space. 
For instance, the system in \cite{brack2020decentralized} suffers from a large storage overhead because it needs to store ciphertexts for all the probable infected users in the system.

\section{Conclusion}
\label{chap:conclusion}

In this paper, we have proposed an efficient and privacy-preserving Blockchain-based infection control system. In our system, a group of health authorities form a consortium Blockchain to trace the close contacts of users who are tested positive for COVID-19. Compared to the existing systems, our system does not depend only on contact tracing to control the infection, but also on limiting the access of public places to non-infected users and recommending the safe-places to visit. We have performed security and privacy analysis to prove the security of our system and its ability to preserve the users' privacy against identification, social graph disclosure, and tracking attacks, while thwarting false reporting attacks. 
We have used the signature aggregation technique and the Bloom filters to enhance the scalability of our system. 
Extensive performance evaluations are conducted and the results demonstrate that the communication, computation, and storage overheads of our system are low. 

\newcolumntype{P}[1]{>{\centering\arraybackslash}p{#1}}
%\balance
\bibliographystyle{IEEEtran}
\bibliography{main}
% \newpage
%\input{Files/bio.tex}

\end{document}